\documentclass[twocolumn]{aastex62}

\usepackage{graphicx}	
\usepackage{amsmath}	
\usepackage{amssymb}	

\graphicspath{{./}{./}}

\received{October 6, 2018}
\submitjournal{ApJ}

\shorttitle{Resolved {[CII]}/FIR deficit in DSFGs}
\shortauthors{Rybak et al.}

\begin{document}

\title{Strong FUV fields drive the [CII]/FIR deficit in $z\sim3$ dusty, star-forming galaxies}

\correspondingauthor{Matus Rybak}
\email{mrybak@strw.leidenuniv.nl}

\author{Matus Rybak}
\affiliation{Leiden Observatory, Leiden University, PO Box 9513, NL-2300 RA Leiden, the Netherlands}
\author{G. Calistro Rivera}
\affiliation{Leiden Observatory, Leiden University, PO Box 9513, NL-2300 RA Leiden, the Netherlands}
\author{J. A. Hodge}
\affiliation{Leiden Observatory, Leiden University, PO Box 9513, NL-2300 RA Leiden, the Netherlands}

\author{Ian Smail}
\affiliation{Centre for Extragalactic Astronomy, Department of Physics, Durham
University, South Road, Durham DH1 3LE, United Kingdom}
\author{F. Walter}
\affiliation{Max Planck Institute for Astronomy, K\"onigstuhl 17, 69117 Heidelberg, Germany}
\author{P. van der Werf}
\affiliation{Leiden Observatory, Leiden University, PO Box 9513, NL-2300 RA Leiden, the Netherlands}
\author{E. da Cunha}
\affiliation{The Australian National University, Mt Stromlo Observatory, Cotter
Rd, Weston Creek, ACT 2611, Australia}
\author{Chian-Chou Chen}
\affiliation{European Southern Observatory, Karl-Schwarzschild-Strasse 2, 85748 Garching bei M\"unchen, Germany}
\author{H. Dannerbauer}
\affiliation{Dpto. Astrof\'isica, Universidad de La Laguna, E-38206 La Laguna, Tenerife, Spain}
\author{R. J. Ivison}
\affiliation{Institute of Astronomy, University of Edinburgh, Royal Observatory, Blackford Hill, Edinburgh EH9 3HJ, United Kingdom}
\affiliation{European Southern Observatory, Karl-Schwarzschild-Strasse 2, 85748 Garching bei M\"unchen, Germany}
\author{A. Karim}
\affiliation{Max-Planck-Institut f\"ur Radioastronomie, Auf dem H\"ugel 69, D-53121 Bonn, Germany}
\author{J. M. Simpson}
\affiliation{Academia Sinica Institute of Astronomy and Astrophysics, No. 1, Sec. 4, Roosevelt Road, Taipei 10617, Taiwan}
\author{A. M. Swinbank}
\affiliation{Centre for Extragalactic Astronomy,Department of Physics,Durham
University, South Road, Durham DH1 3LE, United Kingdom}
\author{J. L. Wardlow}
\affiliation{Physics Department, Lancaster University, Bailrigg, Lancaster LA1 4YB, United Kingdom}

\begin{abstract}

We present 0.15-arcsec (1~kpc) resolution ALMA observations of the [\ion{C}{2}] 157.74~$\mu$m line and rest-frame 160-$\mu$m continuum emission in two $z \sim 3$ dusty, star-forming galaxies - ALESS~49.1 and ALESS~57.1, combined with resolved CO(3--2) observations. In both sources, the [\ion{C}{2}] surface brightness distribution is dominated by a compact core $\leq$1 kpc in radius, a factor of 2--3 smaller than the extent of the CO~(3--2) emission. In ALESS~49.1, we find an additional extended (8-kpc radius), low surface-brightness [\ion{C}{2}] component. Based on an analysis of mock ALMA observations, the [\ion{C}{2}] and 160-$\mu$m continuum surface brightness distributions are inconsistent with a single-Gaussian surface brightness distribution with the same size as the CO(3--2) emission. The [\ion{C}{2}] rotation curves flatten at $\simeq2$~kpc radius, suggesting the kinematics of the central regions are dominated by a baryonic disc. Both galaxies exhibit a strong [\ion{C}{2}]/FIR deficit on 1-kpc scales, with FIR-surface-brightness to [\ion{C}{2}]/FIR slope steeper than in local star-forming galaxies. A comparison of the [\ion{C}{2}]/CO(3--2) observations with PDR models suggests a strong FUV radiation field ($G_0\sim10^4$) and high gas density ($n\mathrm{(H)}~\sim10^4-10^5$~cm$^{-3}$) in the central regions of ALESS~49.1 and 57.1. The most direct interpretation of the pronounced [\ion{C}{2}]/FIR deficit is a thermal saturation of the C$^+$ fine-structure levels at temperatures $\geq500$~K, driven by the strong FUV field.

\end{abstract}


\section{Introduction}
\label{sec:introduction}

Dusty, star-forming, submillimeter galaxies (DSFGs, SMGs) are a major contributor to the global star-formation rate between redshifts $z=2-4$, at an epoch when the star-forming activity of the Universe was at its peak (e.g.,\,\citealt{Casey2014}). Although few in numbers, thanks to their high star-formation rates (SFR $>$100~M$_\odot$ yr$^{-1}$), up to 20\% of all the star formation at $z\sim3$ takes place in SMGs \citep{Swinbank2014}.

The massive dust reservoirs in SMGs absorb the UV/optical radiation from the newborn stars, mostly re-radiating it thermally as a rest-frame FIR/sub-mm continuum\footnote{Following \citep{Casey2014}, we consider SMGs to comprise high-redshift galaxies with a continuum flux $\geq$1~mJy between 250~$\mu$m and 2~mm.}. Therefore, studying the structure and physical properties in these extreme sources requires relying on sub-mm/mm bright tracers -- the dust continuum (which directly traces the obscured star formation) and low-$J$ CO rotational transitions\footnote{In this work, we use the term ``low-$J$'' transitions for the rotational transitions with $J_{upp} \leq 3$.} (which trace the cold, molecular gas that fuels the star formation, \citealt{Carilli2013}).

Besides the FIR continuum and CO emission, the third bright rest-frame FIR tracer of the star-forming interstellar medium (ISM) is the [\ion{C}{2}] 157.74~$\mu$m line, a fine-structure transition of C$^+$ ions. Due to its low ionization energy (11.3~eV) and a relatively low critical density, [\ion{C}{2}] traces of a wide range of ISM phases -- from the ionized \ion{H}{2} regions to warm molecular clouds to diffuse gas. Depending on the environmental conditions, the upper fine-structure level is populated predominantly by collisions with H, H$_2$ or electrons \citep{Goldsmith2012}.

Starting in the early 1990's, systematic studies of [\ion{C}{2}] emission in local galaxies were enabled by the \textit{Infrared Space Observatory} and the Kuiper Airborne Observatory. These observations revealed a tight correlation between the [\ion{C}{2}] line and FIR continuum emission from the heated dust at low SFR surface densities (e.g.,\,\citealt{Stacey1991}). However, this correlation breaks at larger FIR surface brightness $\Sigma_\mathrm{FIR}$ -- the so-called ``\emph{[CII]/FIR deficit}'' (e.g.,\,\citealt{Malhotra1997, Luhman1998, Malhotra2001, Luhman2003}) -- with the [\ion{C}{2}]/FIR ratio decreasing with increasing $\Sigma_\mathrm{FIR}$.

In the last decade, the study of [\ion{C}{2}] emission in the nearby Universe was revolutionized by \textit{Herschel}. The largest sample of [\ion{C}{2}] observations in nearby starburst galaxies was presented by \citet{Diaz2013}, who obtained PACS spectroscopic observations of the 241 galaxies from the Great Observatories All-sky LIRG Survey (GOALS, \citealt{Armus2009}). Further systematic studies of the [\ion{C}{2}] emission in local galaxies have confirmed strong correlation of the [\ion{C}{2}]/FIR deficit with $\Sigma_\mathrm{SFR}$ down to 200-pc scales, in a wide range of environments from normal galaxies \citep{Smith2017, Herrera2018} to starbursts \citep{Diaz2017} and AGN hosts \citep{Herrera2018}.

At the highest redshifts ($z>4$), the importance of the [\ion{C}{2}] line increases dramatically, as the raised CMB temperature renders the low-$J$ CO emission undetectable, while the [\ion{C}{2}] line remains relatively unaffected \citep{daCunha2013, Vallini2015, Lagache2018}. Indeed, with the advent of Atacama Large Millimeter/submillimeter Array (ALMA), [\ion{C}{2}] observations are now increasingly used to determine redshifts and dynamical masses of high-redshift galaxies, including some of the most distant systems \citep{Walter2009, Brisbin2015, Gullberg2015, Oteo2016, Carniani2017, Decarli2018, Smit2018}. Most recently, \citet{Gullberg2018} presented deep, 30-mas resolution (200~pc physical scale) ALMA observations of the [\ion{C}{2}] line in four (unlensed) $z$=4.4--4.8 galaxies. Although their observation suffered from a very sparse $uv$-plane coverage, they found the resolved [\ion{C}{2}]-FIR deficit at $z\sim$4.5 to follow the trend seen in local galaxies. Similarly, high-resolution ALMA observations of [\ion{C}{2}]/FIR deficit in two strongly lensed galaxies at $z$=1.7 and 5.6 were recently presented by \citet{Lamarche2018} and \citet{Litke2018}, respectively, showing a pronounced [\ion{C}{2}]/FIR deficit ($10^{-4} - 10^{-3}$) on (sub)kpc scales.

Despite the recent progress, it is still unclear how well the results and relations derived from local observations hold for the high-redshift population, especially the intensely star-forming high-redshift SMGs, with $\Sigma_\mathrm{SFR}$ two to three orders of magnitude higher than the local star-forming galaxies ($10^{-3}-1$~M$_\odot$ yr$^{-1}$ kpc$^{-2}$, e.g.,\,\citealt{Smith2017}). To directly compare the [\ion{C}{2}] emission in high-redshift SMGs to the local galaxies and study its connection to the star-formation, high-resolution (kpc-scale) observations of the [\ion{C}{2}] emission, alongside the rest-frame FIR continuum (tracing the obscured star formation) and the low-$J$ CO emission tracing the molecular gas are necessary. 

At high redshift, such resolved, multi-tracer studies are limited by the angular extent of the source (few arcseconds at most) and the need for robust redshifts to ensure that both [\ion{C}{2}] and low-$J$ CO emission are observable from the ground. For example, \citet{Stacey2010} compared unresolved [\ion{C}{2}]/FIR/CO($J_{upp}\leq4$) observations of a heterogeneous sample of $z=1-2$ galaxies. \citet{Gullberg2015} compared [\ion{C}{2}], FIR and CO(2--1)/(1--0) observations in 20 strongly lensed SMGs ($z_\mathrm{source}$=2.1--5.7); they found the [\ion{C}{2}] and CO line profiles to be very similar, suggesting that they originate from the same source-plane regions. However, the limited spatial resolution of these observations prevented a robust source-plane reconstruction of the CO/[\ion{C}{2}] emission; the \emph{differential magnification} bias (e.g.,\,\citealt{Serjeant2012}) therefore could not be eliminated. In addition, several individual sources at $z>4.5$ have been studied in both [\ion{C}{2}] and low-$J$ CO emission (e.g.,\,\citealt{Walter2009, Huynh2014, Cicone2015}), though these tend to be extreme sources in terms of FIR brightness and AGN activity. 

Finally, the [\ion{C}{2}] line has been proposed as an alternative to CO emission as a molecular gas tracer (e.g.,\,\citealt{Zanella2018}). However, in SMGs, the spatial extent of CO emission has been shown to vary strongly with $J_\mathrm{upp}$ (e.g.,\,\citealt{Ivison2011,Riechers2011}). If [\ion{C}{2}] emission traces only a subset of the molecular gas reservoir, [\ion{C}{2}]-based mass estimates might be severely biased.

In this paper, we explore a new regime in resolved, multi-tracer studies by investigating resolved [\ion{C}{2}], FIR continuum and CO(3--2) emission on kpc-scales in two (unlensed) $z\sim3$ sources. This allows us to address the following questions:

\begin{itemize}

\item How does the resolved [\ion{C}{2}]/FIR ratio at $z\sim$3 compare to that seen in local and high-redshift star-forming galaxies?
\item What physical mechanism drives the [\ion{C}{2}]/FIR deficit in SMGs?
\item How well does the [\ion{C}{2}] emission trace the molecular gas reservoir in SMGs?

\end{itemize} 

Compared to the high-redshift, high-resolution [\ion{C}{2}]-only studies (e.g.,\,\citealt{Gullberg2018, Zanella2018}), resolved emission line maps of two different species (C$^+$ and $^{12}$CO) allow us to study the relation between [\ion{C}{2}] emission and the colder molecular gas.

This paper is structured as follows: in Section~\ref{sec:observations}, we give the details of ALMA observations. Section~\ref{sec:Results} details the processing of the data in both the image- and $uv$-plane, the assessment of the systematic errors and kinematic modelling. Section~\ref{sec:discussion} presents the spatial and kinematic comparison of the interpretation of the [\ion{C}{2}]/CO(3--2)/FIR observations, results of PDR modelling and a discussion of the physical processes driving the [\ion{C}{2}]/FIR deficit. Finally, Section~\ref{sec:conclusions} summarizes the conclusions of this paper.

Throughout this paper we use a flat $\Lambda$CDM cosmology from \citet{Planck2015}. We adopt the CO(3--2) spectroscopic redshifts from \citet{CR18}: $z$=2.943$\pm$0.001 and $z$=2.943$\pm$0.002 for ALESS 49.1 and 57.1. Consequently, 1~arcsec corresponds to 7.9~kpc for both ALESS~49.1 and ALESS~57.1; the luminosity distance to both sources is 25445~Mpc \citep{Wright2006}.

\section{Observations and data reduction}
\label{sec:observations}

\subsection{Sample selection}
The two galaxies analyzed in this paper were identified by \citet{Hodge2013} as a part of the ALESS survey. The ALESS survey was an ALMA Cycle~0 870~$\mu$m imaging campaign targeting all 126 sources discovered in the LABOCA Extended \textit{Chandra} Deep Field South Submillimeter Survey (LESS, \citealt{Weiss2009}). With ALMA Cycle~0 observations providing a significant improvement over the LABOCA map in both resolution (beam area reduced by a factor of $\sim$200) and sensitivity (increased by a factor of $\sim$3), the ALESS survey identified 99 distinct sub-mm bright galaxies in its primary sample \citep{Hodge2013}. Out of these, at the time of the proposal (2015 April) only four -- ALESS 49.1, 57.1, 67.1 and 122.1 -- had robust redshifts that allowed for ALMA observations of both the low-$J$ CO and [\ion{C}{2}] lines. The spectroscopic redshifts were acquired using VLT-FORS2/VIMOS and Keck-DEIMOS, and are based on multiple line detections \citep{Danielson2017}.

In this paper, we present the ALMA Band~8 observations targeting the [\ion{C}{2}] line and rest-frame 160-$\mu$m continuum. The corresponding Band~3 observations, targeting the CO~(3-2) ($\nu_\mathrm{rest}$=345.795~GHz) emission, were recently presented by \citet{CR18}.

\subsection{ALMA Band~8 observations}
\label{subsec:obs_band8}

The observations were carried out as part of the ALMA Cycle~3 Project \#2015.1.00019.S (PI: J. Hodge) on 2016 August 12. Only ALESS~49.1 and ALESS~57.1 ([\ion{C}{2}] line in ALMA Band~8) were observed; ALESS~67.1 ($z=2.12$) and ALESS~122.1 ($z=2.02$) have the [\ion{C}{2}] line in ALMA Band~9, and were not observed. The total time including calibration and overheads was 72~mins, with an on-source time of 11~mins per target. The array configuration consisted of 38 12-m antennas, with baselines extending up to 1400~m. The largest angular scale\footnote{The largest angular scale is estimated as 0.983/$\lambda_{5\%}$, where $\lambda_{5\%}$ is the 5-th percentile $uv$-distance (ALMA Cycle~5 Technical Handbook).} of the observations is $\sim$1.9~arcsec for both sources. The primary beam FWHM is 14.1~arcsec. Synthesized beam sizes and $\sigma_{rms}$ for the 160-$\mu$m continuum and the [\ion{C}{2}] emission are listed in Table~\ref{tab:targets+observed}. The target elevation range was 66--73~deg for ALESS 49.1 and 67--76~deg for ALESS~57.1.

The frequency setup was configured in four spectral windows (SPWs) in Band~8. The individual SPWs were centered at 481.953, 483.183, 493.506, 495.386 GHz. Each SPW was split into 480 frequency channels 3.906~MHz wide, giving a total bandwidth of 1.875~GHz per SPW. The radio-velocity resolution was 2.42~km~s$^{-1}$. Both the Stokes $XX$ and $YY$ parameters were observed.

The data were calibrated using the standard ALMA pipeline, with additional flagging necessary to remove atmospheric features (see Section~\ref{subsec:imaging}). All the visibility data processing apart from imaging was performed using {\sc Casa} versions 4.7 and 5.0 \citep{McMullin2007}.

The spectral structure of the [\ion{C}{2}] line overlaps with atmospheric absorption features at 481.2 and 481.6~GHz. Consequently, the noise level in the affected channels is raised by a factor of $\sim$2. This issue is particularly severe for ALESS~49.1. Another atmospheric feature is located in a line-free SPW at 496.35 GHz; here, the affected channels were flagged to improve the continuum SNR. As will be outlined in Section~\ref{subsec:imaging}, we use a channel-dependent threshold for the deconvolution process to avoid introducing noise features from the affected channels.

To increase the signal-to-noise ratio, each line was split into frequency bins $\sim$120 km~s$^{-1}$ wide. Additionally, we time-averaged the data using a 30-second bin; this corresponds to an average intensity loss of $<$0.5~per~cent at 5~arcsec from the phase-tracking centre, which we consider negligible\footnote{\citet{SynthesisImaging}, equation (18--42).}. The time-bin size was chosen so as to prevent significant time-averaging smearing. The two linear polarizations were combined into the Stokes intensity $I$.

\subsection{ALMA Band 3 observations}

ALMA Band~3 observations of the CO(3--2) and rest-frame 1.0-mm continuum in ALESS~49.1 and 57.1 were presented by \citet{CR18}. These consisted of Cycle~2 observations of ALESS 49.1 and 57.1 (Project \#2013.1.00470.S; PI: J. Hodge) at 0.34-0.67~arcsec resolution, and additional Cycle~4 observations of ALESS 49.1 (Project \#2016.1.00754.S; PI: J. Wardlow) at 1.1-arcsec resolution. The naturally-weighted Band 3 synthesized beam size is 0.69$\times$0.63 arcsec for ALESS 49.1 (after concatenating the Cycles 2 and 4 data) and 0.67$\times$0.60~arcsec for ALESS 57.1, with a continuum $\sigma_{rms}$ of 17.6 and 19.5~$\mu$Jy beam$^{-1}$, respectively. For a detailed description of the data and the resulting analysis, we refer the reader to \citet{CR18}.

\section{Results}
\label{sec:Results}

\begin{table*}
\caption{Source positions (corresponding to the 160-$\mu$m continuum surface brightness maximum), redshifts, ALMA Band~8 naturally-weighted beam sizes and noise levels, observed 160-$\mu$m continuum/[\ion{C}{2}] flux-densities and the FWHM of the [\ion{C}{2}] and CO(3--2) lines. The redshifts are based on CO(3--2) observations \citet{CR18}. The spatially-integrated 160-$\mu$m and [\ion{C}{2}] flux-densities are measured within a circular aperture with a 1.0-arcsec diameter, centered on the continuum surface brightness maximum. \label{tab:targets+observed}}

\begin{center}
 \begin{tabular}{@{}ll|cc @{}}
 \hline \hline
Source & & ALESS 49.1 & ALESS 57.1 \\
\hline
RA (J2000) & & 3:31:24.71 & 3:31:51.94 \\
DEC (J2000) & & $-$27:50:46.9 & $-$27:53:27.0 \\
$z\,^a$ & & 2.943$\pm$0.001 & 2.943$\pm$0.002 \\
Beam FWHM & [arcsec] & 0.16$\times$0.12 & 0.16$\times$0.12 \\
Beam PA & [deg] & 53 & 56 \\
 \hline

$S_\mathrm{160 \mu m}$ & [mJy] & 10.7$\pm$1.0 & 8.2$\pm$0.5 \\
$I_\mathrm{160 \mu m}^{max}$ & [mJy beam$^{-1}$] & 3.42$\pm$0.16 & 1.53$\pm$0.17 \\

$S_\mathrm{[CII]}\,^b$ & [mJy] & 15.4$\pm$2.8 & 8.8$\pm$2.5 \\
$I^\mathrm{max}_\mathrm{[CII]}\,^b$ & [mJy beam$^{-1}$] & 2.0$\pm$0.4 & 3.0$\pm$0.4 \\
FWHM$_\mathrm{[CII]}$ & [km s$^{-1}$] & 600$\pm$130 & 390$\pm$70 \\ 
FWHM$_\mathrm{CO}\,^a$ & [km s$^{-1}$] & 610$\pm$30 & 360$\pm$90 \\ 
\hline

\multicolumn{4}{l}{$^a$ adopted from \citet{CR18}.}\\
\multicolumn{4}{l}{$^b$ integrated over 800 and 710km s$^{-1}$ for ALESS~49.1 and 57.1, respectively.}\\

 \end{tabular}
 \end{center}

\end{table*}

\subsection{Image analysis}
\label{subsec:imaging}
\subsubsection{Imaging}
We perform synthesis imaging of the visibility data using the {\sc Ws-Clean} algorithm introduced by \citet{Offringa2014}, specifically its multi-scale version \citep{Offringa2017}. The multi-scale {\sc Ws-Clean} is an advanced deconvolution algorithm with a multi-scale, multi-frequency capability \citep{Offringa2017}. Another advantage of the {\sc Ws-Clean} as opposed to the {\sc Casa} implementation is the channel-dependent deconvolution threshold. As the noise level changes appreciably with frequency due to atmospheric lines, this prevents us from introducing noise-peaks from the affected channels into the reconstructed images.

For the line imaging, we first subtract the continuum by linearly interpolating the line-free channels in SPWs 1, 2, and 3 and subtract the continuum slope from the line-containing channels. The continuum channels overlapping with the atmospheric lines were flagged before the continuum subtraction. For the continuum imaging, we discard the entire SPW 0 and the line-containing channels in SPW1, as well as the channels affected by the atmospheric feature around 496.35 GHz.

The data were deconvolved on a sky-plane grid of 1024 $\times$ 1024 5-mas pixels (total FoV size = 5.115$\times$5.115 arcsec), using natural weighting. We use the automatic SNR-based masking, with \texttt{auto-mask} SNR threshold of 2. 

For consistency, we re-image the CO(3--2) data of \citet{CR18} using the exactly same procedure as for the [\ion{C}{2}] data; these result in minor ($\leq$10\%) changes in the rms noise and the inferred CO(3--2) luminosity. We use these re-imaged data only for the spectral comparison in Section \ref{fig:moments+spectra}; the CO(3--2) source size and hence the bulk of our analysis in Section~\ref{sec:discussion} is based on the $uv$-plane analysis and hence is unaffected by the imaging procedure. We adopt \citet{CR18} CO(3--2) luminosity and gas mass estimates for the remainder of this paper.

\subsubsection{ALMA 160-$\mu$m continuum}

The rest-frame 160-$\mu$m continuum is detected in both sources at $>$10~$\sigma$-confidence (Figure~\ref{fig:aless_images}). In both sources, the continuum emission shows only a single brightness peak at 0.15-arcsec resolution. In ALESS~49.1, the continuum emission is almost circularly symmetric, with a minor extension in the east-west direction. In ALESS~57.1, the continuum is noticeably extended in the east-west direction (axis ratio $\sim$2:1). The spatially-integrated 160-$\mu$m continuum flux-density for ALESS~49.1 and 57.1 is given in Table~\ref{tab:targets+observed}. The 160-$\mu$m flux-density is calculated from an aperture with a 1.0~arcsec diameter, centered on the 160-$\mu$m continuum surface brightness maximum. The aperture size was chosen based on the FIR continuum and [\ion{C}{1}] sizes determined from the $uv$-plane fitting (Section~\ref{subsec:uv-plane} as we do not expect significant surface brightness contribution $\geq$0.5~arcsec from the centre of the source.

\subsubsection{[CII] emission}

\begin{figure*}
\begin{centering}
\includegraphics[height=4.5cm, clip=true]{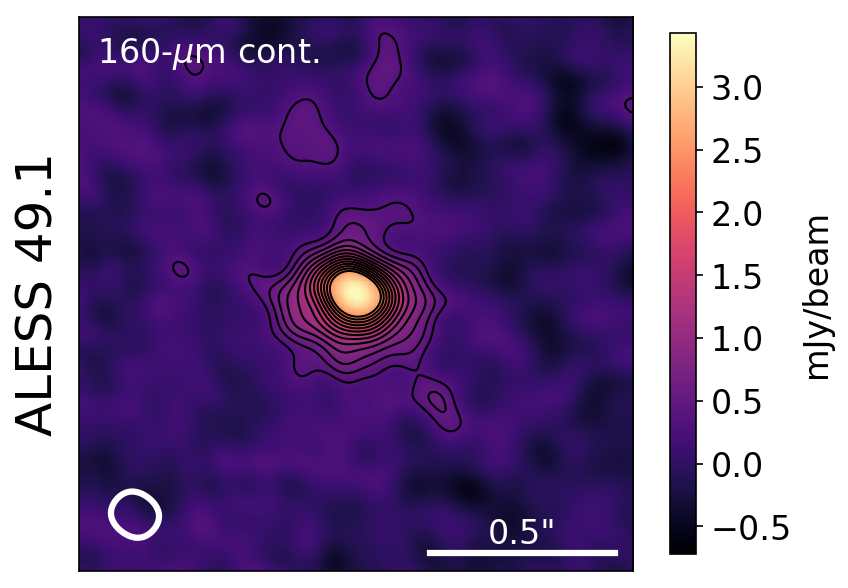}
\includegraphics[height=4.5cm, clip=true]{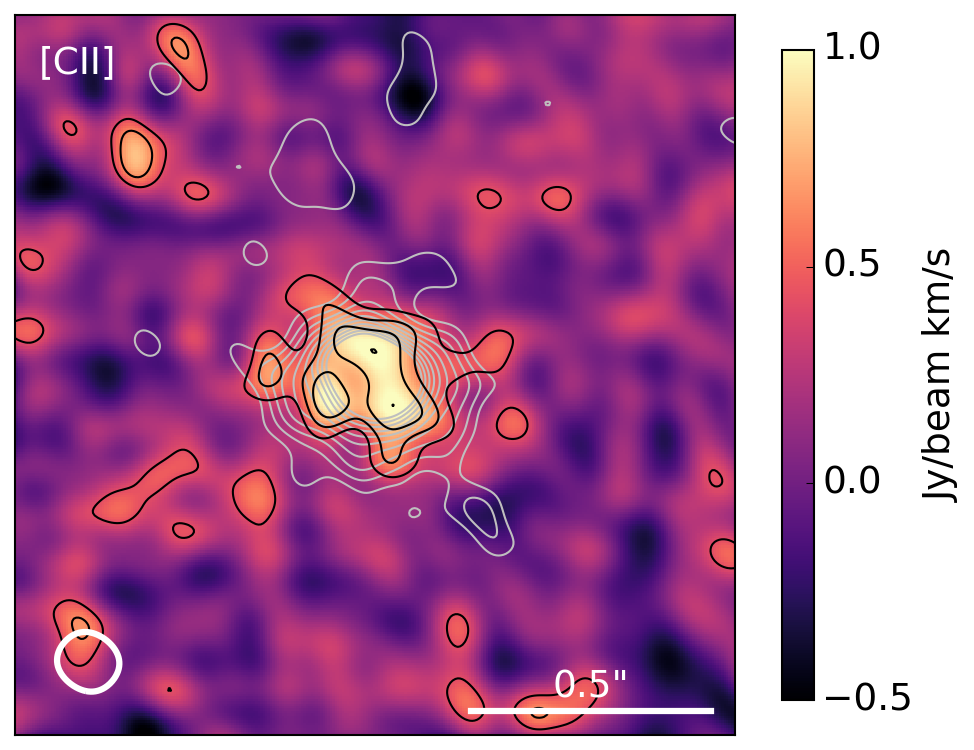}

\includegraphics[height=4.5cm, clip=true]{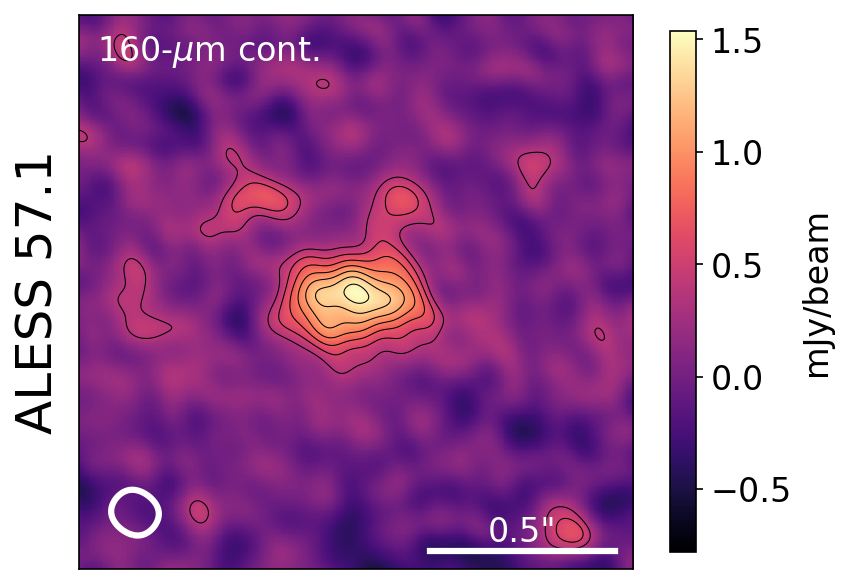}
\includegraphics[height=4.5cm, clip=true]{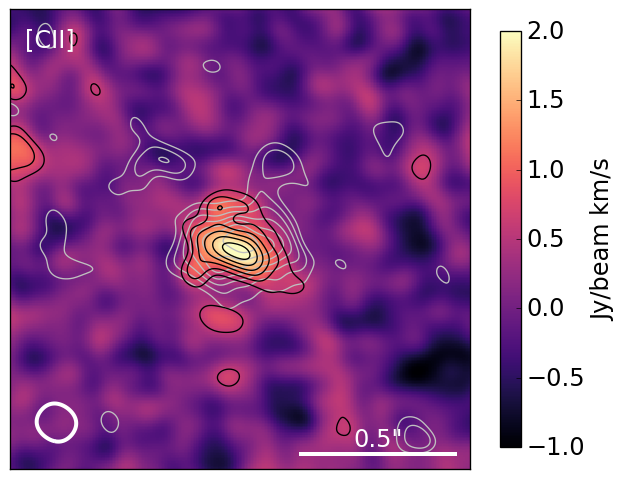}\\

\caption{ALESS 49.1 and 57.1 imaging: continuum (right) and integrated [\ion{C}{2}] emission (left) maps with continuum contours overplotted in grey.
The contours start at 2$\sigma$ level and increase by 1$\sigma$. The FWHM beam size is indicated by the ellipse in the lower left corner. The [\ion{C}{2}] and 160-$\mu$m emission is well-resolved in both sources, and almost co-spatial. \label{fig:aless_images}}
\end{centering}
\end{figure*}

The velocity-integrated maps of the [\ion{C}{2}] emission are presented in Figure~\ref{fig:aless_images}. The [\ion{C}{2}] emission is detected at 5- and 8$\sigma$ significance in ALESS~49.1 and 57.1, respectively. The [\ion{C}{2}] emission is relatively compact ($<$0.5 arcsec diameter) in both ALESS~49.1 and 57.1, similar in extent to the 160-$\mu$m continuum. The [\ion{C}{2}] emission in ALESS 57.1 is highly elliptical (axis ratio $\geq$2:1) and elongated in the east-west direction; the [\ion{C}{2}] does not show any significant offset from the 160-$\mu$m peak. Note that the low-significance clumpy substructure such as that seen in ALESS~49.1 [\ion{C}{2}] maps is often an artifact of low SNR (e.g.,\,\citealt{Hodge2016, Gullberg2018}), rather than a real physical feature.

\subsection{\textit{uv}-plane analysis}
\label{subsec:uv-plane}

We estimate the size of the continuum and [\ion{C}{2}] emission regions by directly fitting the observed visibility function. Given the relatively low SNR of the data, we assume the surface brightness distribution to follow a circularly symmetric Gaussian profile.

The size is measured using the spectrally-averaged continuum and line datasets. To correct the offset between the phase-tracking centre and the centroid of the surface brightness distribution given in Table~\ref{tab:targets+observed}, each dataset is phase-shifted to center the field-of-view on the centroid of the source. The data are then radially binned into bins of equal width. To test the robustness of the $uv$-plane fitting against the $uv$-bin size, we vary the $uv$-bin size from 5 to 50~k$\lambda$.

\begin{figure*}
\begin{centering}
\includegraphics[height=6.5cm, clip=true]{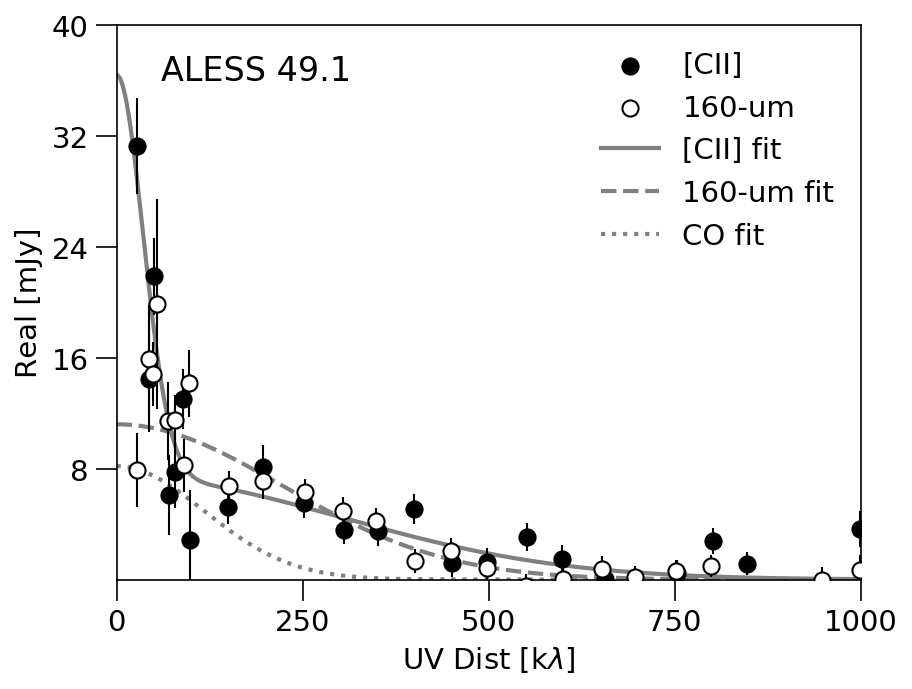}
\includegraphics[height=6.5cm, clip=true]{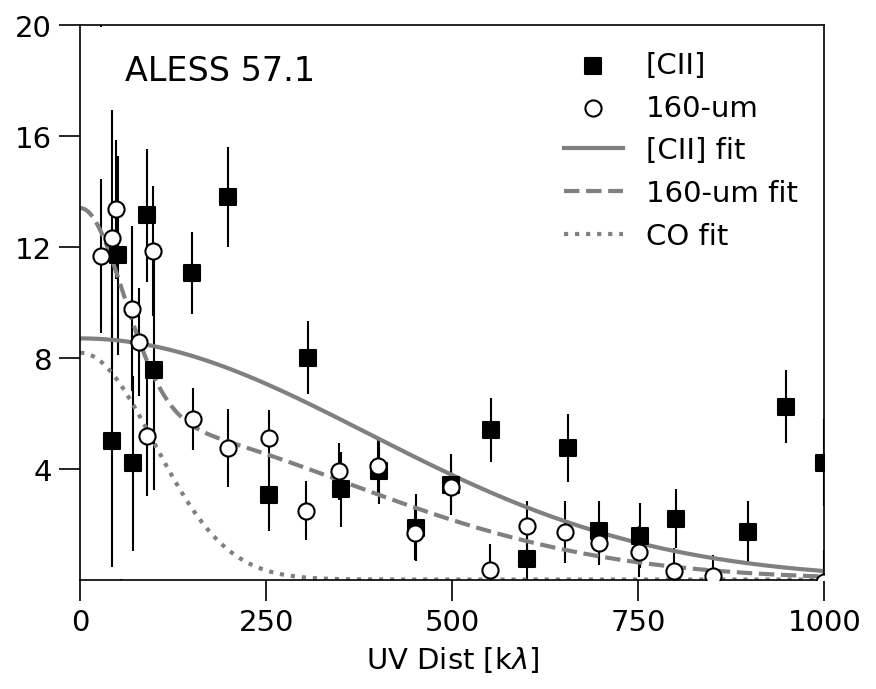}

\caption{Azimuthally averaged visibility function for ALESS 49.1 (left) and ALESS 57.1 (right), for the [\ion{C}{2}], 160-$\mu$m continuum with best-fitting model (full and dashed lines), and best-fitting single-Gaussian profiles to the CO(3--2) data (dotted line, increased by a factor of 10, \citealt{CR18}). For added clarity, the data is binned with a 10 -- 50~k$\lambda$ bin size and truncated at 1000~k$\lambda$. In ALESS~49.1, the excess [\ion{C}{2}] flux at short baselines - corresponding to the extended component - is clearly visible. \label{fig:aless_uvplots} }
\end{centering}
\end{figure*}

We fit each $uv$-plane dataset with (1) a single Gaussian profile, (2) two Gaussian components to investigate the possibility of having compact, bright [\ion{C}{2}] emission embedded in an extended, low surface-brightness component; (3) a combination of a single Gaussian profile and a constant term, corresponding to a point-source in the image plane. To determine whether the two-component model significantly improves the goodness-of-fit compared to the one-component model, we compare the two models using the F-test \citep{Bevington1969}. The single-component model is preferred for the continuum emission in ALESS~49.1 and [\ion{C}{2}] emission in ALESS~57.1, independently of the $uv$-bin size. We will address the robustness of the inferred $R_{1/2}$ in Section~\ref{subsec:mock_data}. The two-component model is strongly ($p>0.95$) preferred for the [\ion{C}{2}] emission in ALESS~49.1 and the continuum emission in ALESS~57.1. We do not find the Gaussian $+$ constant-term (point-source) model to be preferred over the single-Gaussian model for any dataset considered. The best-fitting values for the two-component model are listed in Table~\ref{tab:source_properties}. Figure~\ref{fig:aless_uvplots} shows the visibility function for the [\ion{C}{2}] and 160-$\mu$m continuum, as well as the CO(3--2) observations from \citet{CR18}, and the corresponding best-fitting profiles.

In physical units, for the single-Gaussian models, the [\ion{C}{2}] and 160-$\mu$m emission are rather compact (0.8--1.4~kpc) in both sources. The [\ion{C}{2}] and 160-$\mu$m continuum sizes from the single-Gaussian fitting agree within 1--2$\sigma$ uncertainty. For the two-Gaussian models, the compact and extended [\ion{C}{2}] components in ALESS~49.1 have half-light radii of 1.01$\pm$0.12~kpc and 8.7$\pm$1.6~kpc, respectively. The compact and extended 160-$\mu$m components in ALESS~57.1 have half-light radii of 0.86$\pm$0.07 and 5.3$\pm$1~kpc, respectively.

For the two-component [\ion{C}{2}] model in ALESS~49.1, the extended [\ion{C}{2}] components accounts for up to 80\% of the total [\ion{C}{2}] luminosity. Note that the systematic uncertainty on this estimate might be significant, as it is unclear how much does the extended component depart from the assumption of a circular Gaussian profile. For the [\ion{C}{2}] emission in ALESS~57.1, the single-Gaussian model is preferred. However, if we speculate that ALESS~57.1 has an extended [\ion{C}{2}] component with the same size as in ALESS~49.1 ($R_\mathrm{1/2}^\mathrm{ext}=1.01$~arcsec), the 3$\sigma$ upper limit on the total flux-density contributed by this hypothetical component is 80\%. Similarly, if we speculate that the 160-$\mu$m continuum in ALESS~49.1 has an extended component identical in size to that in ALESS~57.1 ($R_\mathrm{1/2}^\mathrm{ext}=0.67$~arcsec), the 3$\sigma$ upper limit on the flux contained in this hypothetical component is 50\%.

How will the extended components in ALESS~49.1 [\ion{C}{2}] emission and ALESS~57.1 160-$\mu$m continuum contribute to the observed [\ion{C}{2}]/160-$\mu$m surface brightness distribution? The half-light radius of the ALESS~57.1 160-$\mu$m extended component is well below the maximum recoverable scale (1.9~arcsec) and therefore should be fully accounted for in the synthesized images. This is supported by the results from the spectral energy distribution modeling (Section~\ref{subsec:SED_models}), which indicate that the 160-$\mu$m continuum is not significantly resolved out. For the [\ion{C}{2}] emission, the half-light radius of the extended component in ALESS~49.1 is comparable to the maximum recoverable scale and some emission is likely resolved out in the synthesized images. However, while the extended [\ion{C}{2}] component in ALESS~49.1 dominates the total [\ion{C}{2}] luminosity, it contributes only between 5--20\% of the surface brightness across the inner $R<2$~kpc region. It is for this reason that our analysis in Sections 4.2, 4.3 and 4.4 focuses on the central regions ($R\leq2$~kpc) of ALESS~49.1 and 57.1, including the uncertainty from the extended components in further analysis.

\subsubsection{How robust are the source sizes determined from the uv-plane fitting?}
\label{subsec:mock_data}

Before analyzing our resolved [\ion{C}{2}] and 160-$\mu$m continuum observations, we assess the reliability of the inferred source sizes. In particular, we investigate a possibility that the [\ion{C}{2}]/160-$\mu$m emissions follow CO(3--2) surface distribution, and that the source sizes inferred in Section~\ref{subsec:uv-plane} are an artifact of the limited short-spacing coverage of our Band~8 observations. We estimate this uncertainty using simulated ALMA observations.

The mock observations are created as follows. First, we calculate the mean and rms scatter for the [\ion{C}{2}] and 160-$\mu$m continuum visibilities within a given $uv$-bin. We then subtract the mean signal from the data, which gives us a $uv$-plane coverage corresponding exactly to a given observation, along with a realistic noise measurement for each baseline. We choose this approach to account for the different noise levels for each dataset, and the effect of the atmospheric lines on our [\ion{C}{2}] data. We then inject an artificial source into the data, generating 1000 datasets with different noise realizations for each source. We consider sources with a half-light radius of 0.1--1.0 arcsec and peak surface brightness of 0.05 -- 4.0~mJy beam$^{-1}$. Finally, we bin the mock visibilities in the $uv$-plane using exactly the same procedure as applied to real data in Section~\ref{subsec:uv-plane}.

\begin{figure*}
\begin{centering}
\includegraphics[width=8.0cm, clip=true]{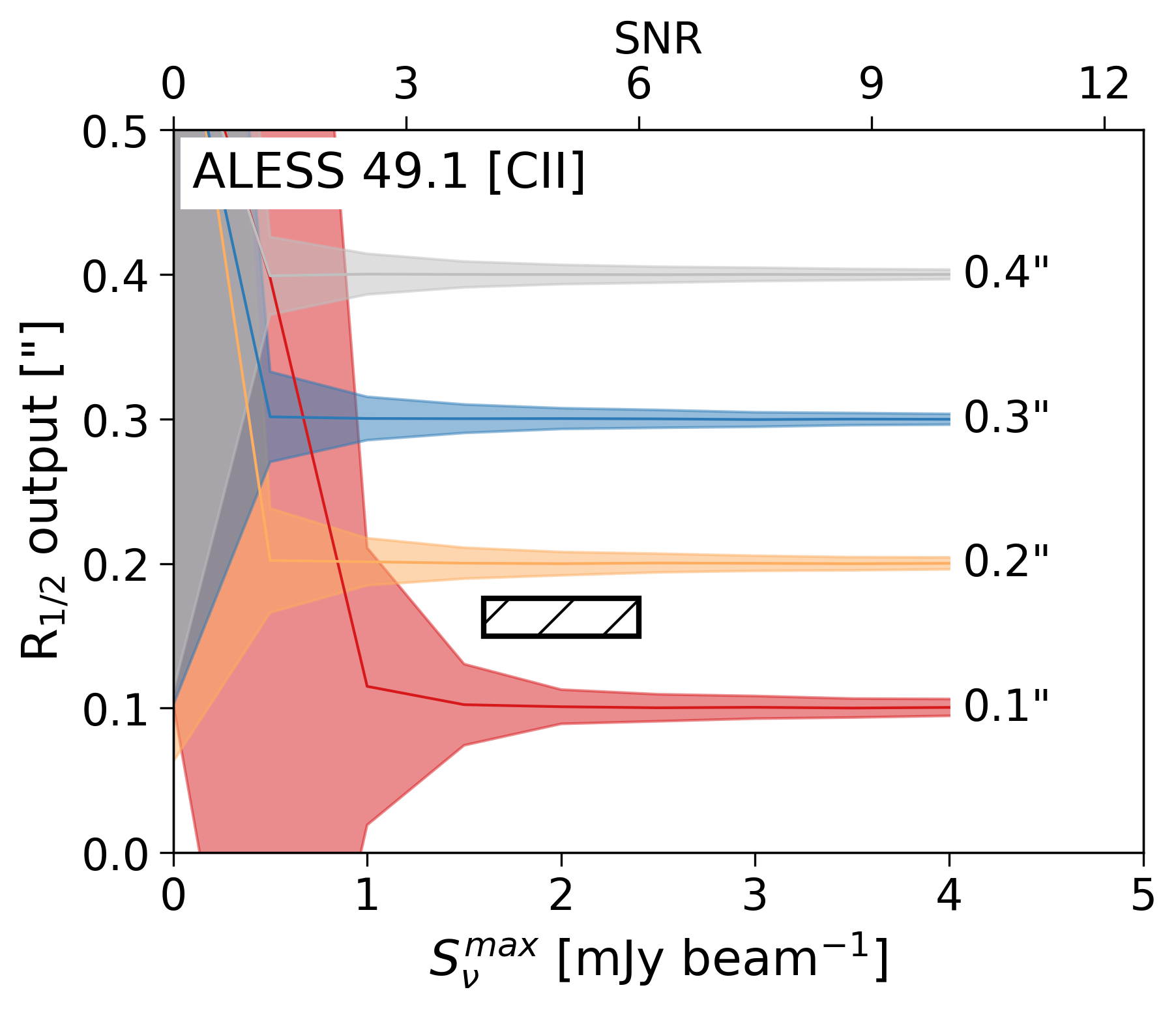}
\includegraphics[width=8.0cm, clip=true]{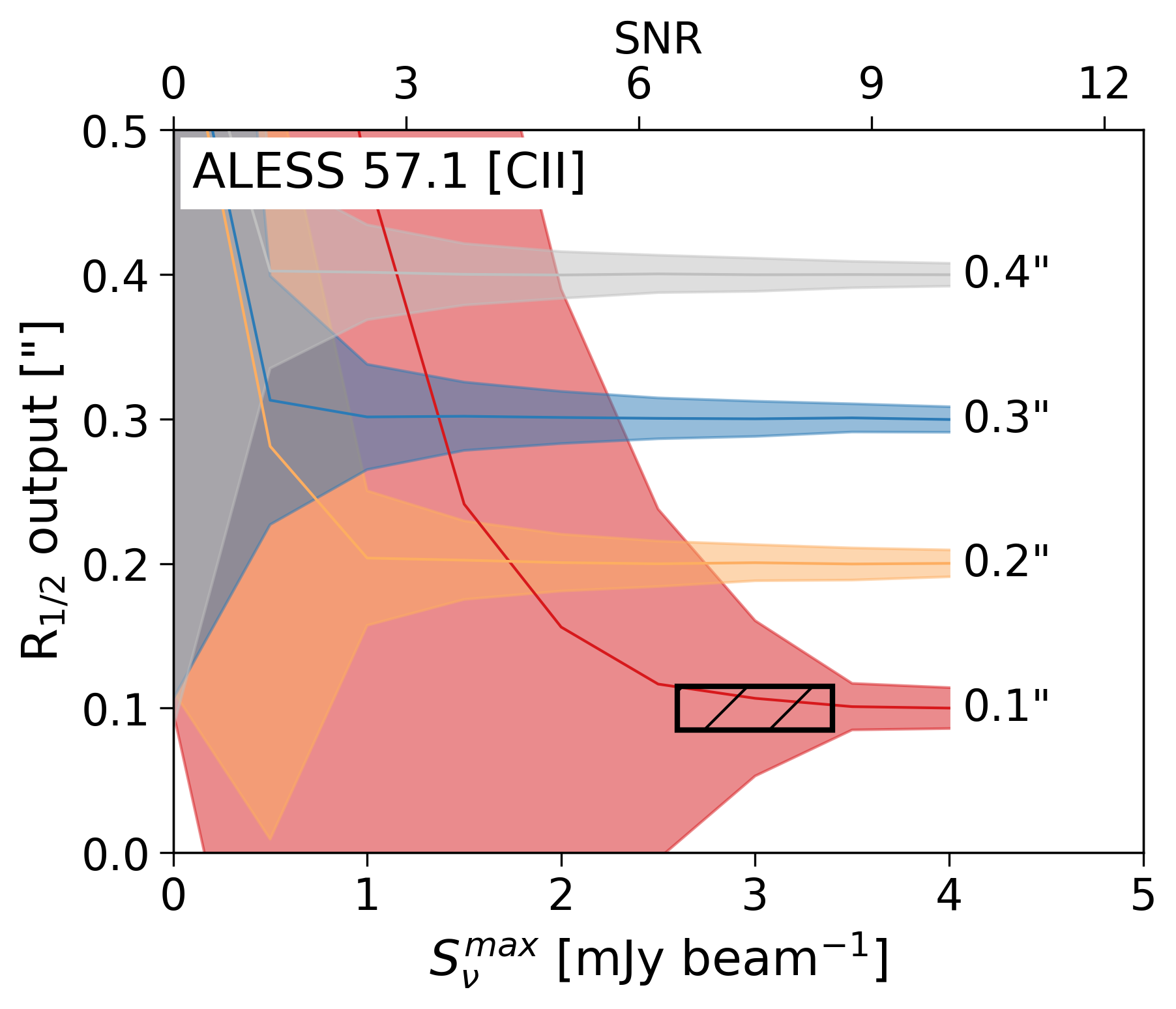} \\
\vspace{0.5cm}
\includegraphics[width=8.0cm, clip=true]{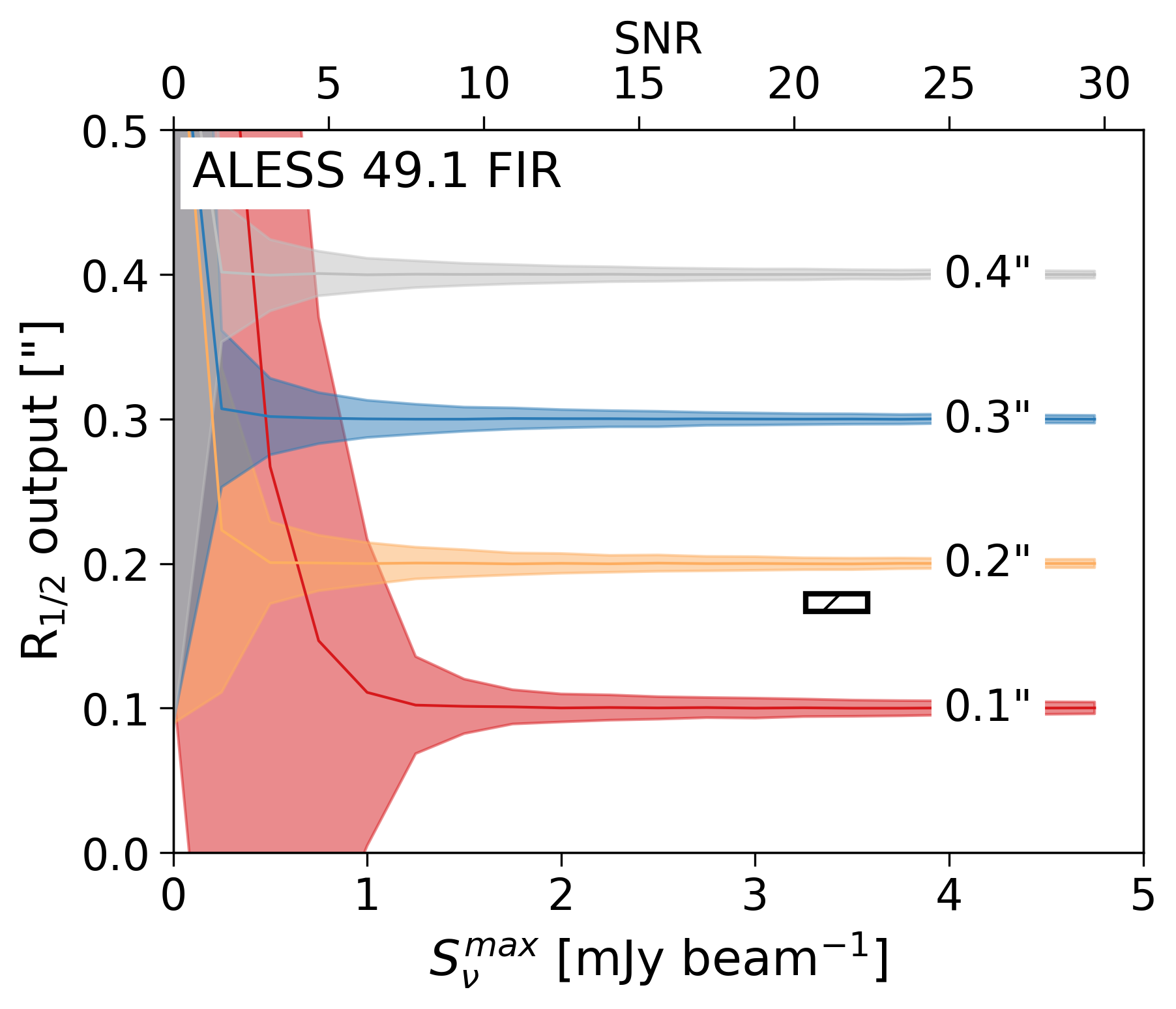}
\includegraphics[width=8.0cm, clip=true]{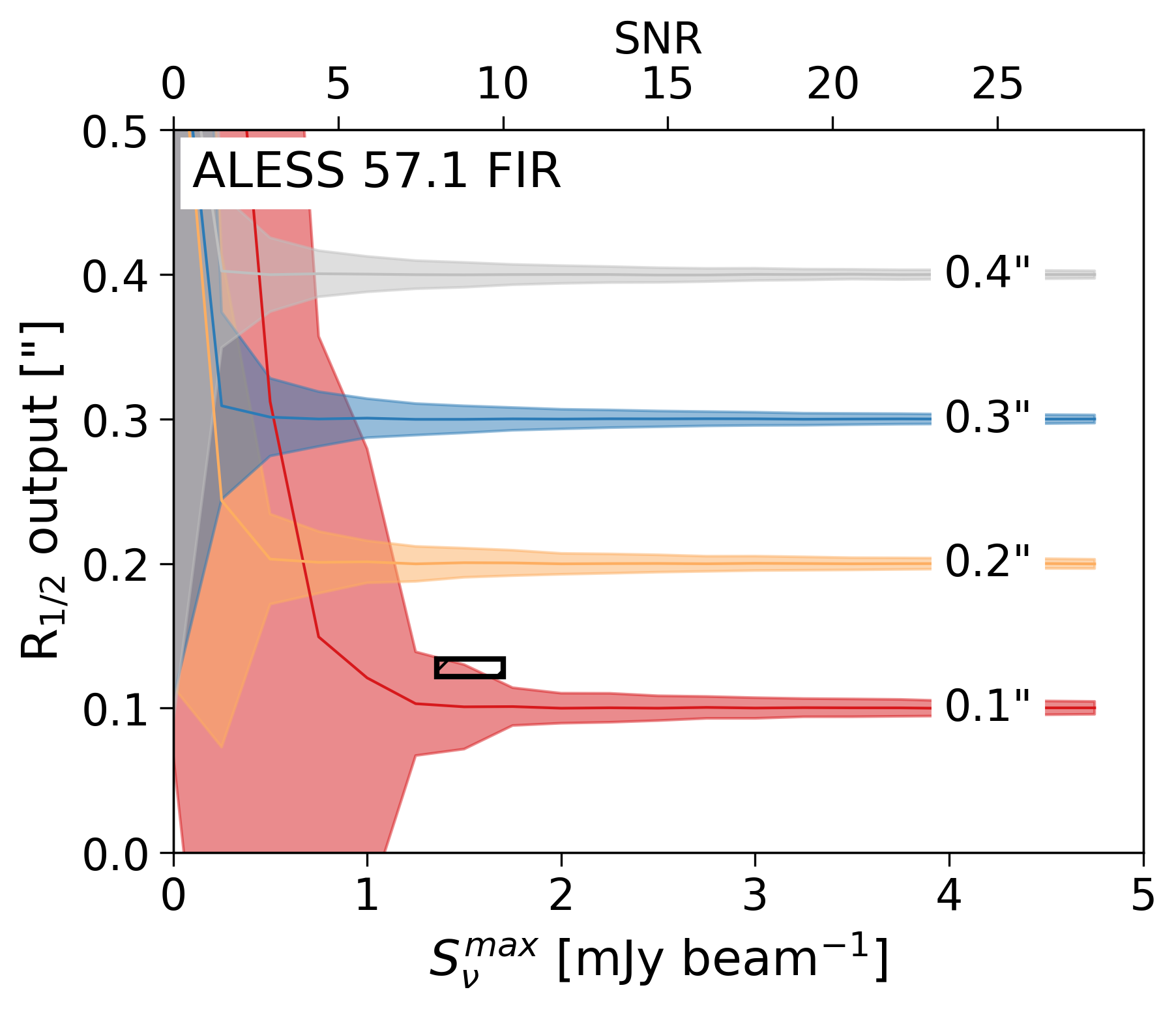}\\

\caption{Comparison of the inferred and true $R_{1/2}$ for the [\ion{C}{2}] line (upper) and 160-$\mu$m data continuum (lower). The bold lines denote the mean inferred $R_{1/2}$ for input $R_{1/2}$=0.1--0.4~arcsec, with coloured regions denoting the 1$\sigma$ uncertainty from the $uv$-plane fitting only. For $R_{1/2}>$0.4 arcsec, the source sizes are inferred robustly and are not shown for clarity. The boxed regions indicate the measured source sizes and peak surface brightness for ALESS~49.1 and 57.1 with 1~$\sigma$ uncertainties. Given the SNR of our observations, we rule out the possibility that [\ion{C}{2}] and 160-$\mu$m continuum follow the same single Gaussian profile as the CO(3--2) emission. \label{fig:mock_radius_1comp}}
\end{centering}
\end{figure*}

Figure~\ref{fig:mock_radius_1comp} shows the inferred radius as a function of the surface brightness maximum alongside the 1$\sigma$ uncertainty from the $uv$-plane fitting. The measured [\ion{C}{2}] and 160-$\mu$m continuum sizes in ALESS~49.1 and 57.1 are all smaller than sizes inferred for input sources with $R_{1/2}\geq0.2$~arcsec in the relevant peak surface brightness range. In other words, given the observed peak surface brightness, $R_{1/2}^\mathrm{[CII]}>$0.2~arcsec source sizes would be recovered within 10\% uncertainty for both ALESS~49.1 and 57.1. For comparison, the CO(3--2) half-light radii in ALESS~49.1 and 57.1 are $R_{1/2}^\mathrm{CO}$=0.33$\pm$0.06 and 0.39$\pm$0.06~arcsec, respectively. 

Therefore, we consider it unlikely that the [\ion{C}{2}]/160-$\mu$m continuum follow a single-Gaussian surface brightness distribution with the same size as CO(3--2) emission (0.33$\pm$0.05 and 0.39$\pm$0.06 arcsec for ALESS~49.1 and 57.1, respectively). However, we can not exclude a combination of a bright compact and faint extended [\ion{C}{2}] and continuum components. In the following analysis, we will focus on the center of the sources where the extended [\ion{C}{2}] component is not expected to contribute significantly.

\subsection{[CII] spectra and kinematics}
\label{subsec:spectra+kinematics}

\begin{figure*}
\begin{centering}
\includegraphics[width=16cm, clip=true]{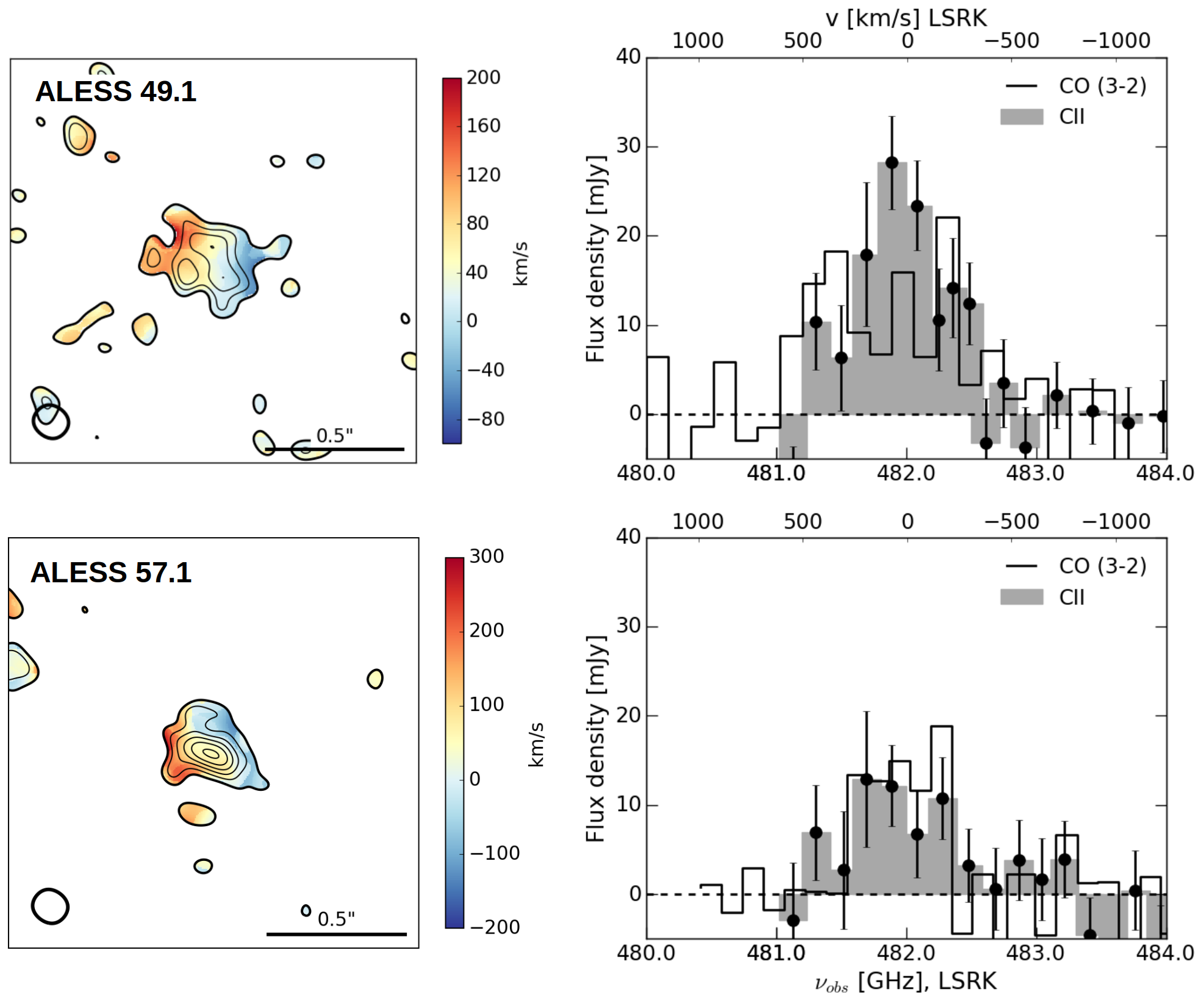}\\

\caption{Moment-one (velocity) maps and [\ion{C}{2}] line profiles for ALESS~49.1 and ALESS~57.1. The contours correspond to [\ion{C}{2}] moment-zero maps as shown in Figure~\ref{fig:aless_images}. The [\ion{C}{2}] velocity fields in both galaxies show a clear gradient across the source, without any major perturbations. The CO(3--2) spectra are taken from \citet{CR18}; the CO(3--2) flux-density is scaled up by a factor of 10 for clarity. The velocity is given in the LSRK frame with respect to the redshift of $z=2.943$, using the radio definition of velocity. The line intensity was measured from images obtained using natural weighting, for an aperture with 1-arcsec / 0.5-arcsec diameter for CO(3--2) and [\ion{C}{2}] emission, respectively, centered on the surface-brightness maximum of the 160-$\mu$m continuum. The error-bars indicate 1$\sigma$ uncertainty. The frequency coverage of our Band~8 observations does not extend below 481.0~GHz. The [\ion{C}{2}] and CO(3--2) line widths match each other (Table~\ref{tab:targets+observed}). \label{fig:moments+spectra}}
\end{centering}
\end{figure*}

Figure~\ref{fig:moments+spectra} presents the [\ion{C}{2}] moment-one (intensity-weighted velocity) maps and the comparison of [\ion{C}{2}]/CO(3-2) line profiles in ALESS~49.1 and 57.1. The moment-one maps reveal a clear velocity gradient across both ALESS~49.1 and 57.1. The spectra were extracted from the naturally-weighted channel maps, using an aperture 1~arcsec ($\sim$8~kpc) in radius for CO(3--2) and 0.5~arcsec ($\sim$4~kpc) in radius for [\ion{C}{2}], given the compact size of the [\ion{C}{2}] emission.

The [\ion{C}{2}] line profile in ALESS~49.1 largely traces the CO~(3--2) profile, exhibiting an increased brightness in the blue channels. In ALESS~49.1, we find a tentative (2.5--3$\sigma$) increase in the [\ion{C}{2}]/CO(3--2) ratio between the centre ($\pm$200~km~s$^{-1}$) and the wings ($\pm(200-600$) km~s$^{-1}$) of the lines. This might be due to: (1) a significant fraction of the [\ion{C}{2}] emission in the reddest and bluest channels being very extended and thus resolved out by our Band~8 imaging, or (2) a spatial variation in the gas conditions. The [\ion{C}{2}/CO(3--2) ratio in ALESS~57.1 is consistent with being constant across the full velocity range.

We model the velocity-fields using the {\sc GalPak3D} software \citep{Bouche2015}. {\sc GalPak3D} uses a Monte-Carlo approach to extract the kinematic and morphological parameters from three-dimensional image cubes, accounting for both the spatial and spectral response of the instrument, assuming a parametric model for a rotating disc. For our simulations, we assume an exponential disc profile - an appropriate choice for ALESS SMGs, which show a mean S\'ersic index of $n=0.9\pm0.2$ (\citealt{Hodge2016}, $n=1$ corresponds to an exponential profile.).
To improve the SNR and the speed of the calculations, we re-sampled our data onto cubes with a pixel size of 25~mas, using natural weighting.

\begin{figure*}
\begin{centering}
\includegraphics[width=18.0cm, clip=true]{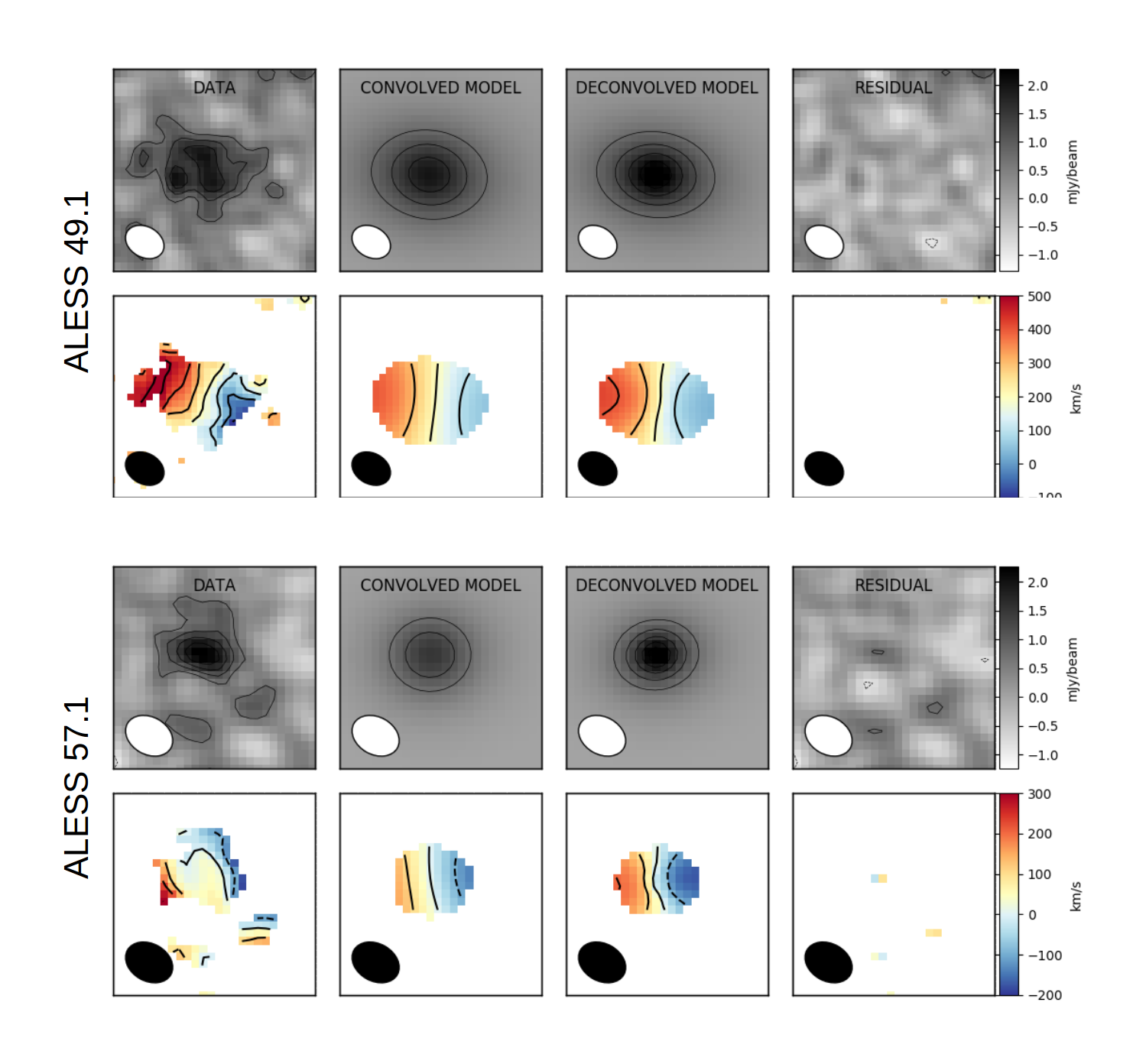}\\

\caption{{\sc GalPak3D} models of the [\ion{C}{2}] velocity field in ALESS 49.1 (upper) and ALESS~57.1 (lower), using an exponential disc model. 
The upper row shows, respectively, the velocity moment-0 (surface brightness) data, best moment-0 model convolved with the beam, best moment-0 model (deconvolved) and the moment-0 residuals. The moment-zero contours start at $\pm2\sigma_\mathrm{rms}$ level and increase in steps of 1$\sigma_\mathrm{rms}$. The moment-one contours are drawn in steps of $\pm$100~km s$^{-1}$. The lower row shows the reconstructed moment-one (velocity) maps for regions with surface brightness $>2\sigma_\mathrm{rms}$. The velocity maps are consistent with the exponential disc model, with no residuals at $>2\sigma_\mathrm{rms}$ significance. The field-of-view size is 1.6$\times$1.6~arcsec and 1.4$\times$1.4 arcsec, respectively. \label{fig:galpak} }

\end{centering}
\end{figure*}

Figure~\ref{fig:galpak} shows the input and reconstructed moment-0 and moment-1 maps for ALESS~49.1 and 57.1. At the SNR of our Band~8 observations, velocity fields in both sources are consistent with an ordered, disc-like rotation. The deconvolved [\ion{C}{2}] rotational curves are shown in Figure\,\ref{fig:rot_curves}, which also lists the FWHM velocity measurements obtained from the fitting of the spatially-integrated CO(3--2) spectra \citep{CR18}. The source inclinations are inferred by fitting the moment-0 map and agree with those derived from CO(3--2) imaging by \citet{CR18} within 1--2$\sigma$ uncertainty.

\begin{table*}
\caption{Inferred source properties for ALESS 49.1 and ALESS 57.1. For the global source properties derived from the SED-fitting (Section~\ref{subsec:SED_models}), we provide the median value of the posterior probability density function. The [\ion{C}{2}] and CO(3--2) line luminosities are spectrally integrated over the full extent of the [\ion{C}{2}], rather than over FWHM only (as in \citealt{CR18}) as the [\ion{C}{2}] line exhibits a non-Gaussian profile. The source sizes are inferred from the $uv$-plane fitting (Section~\ref{subsec:uv-plane}); we list the best-fitting one-component Gaussian model parameters, as well as the two-component Gaussian models if preferred by evidence. The source inclinations and [\ion{C}{2}]-based dynamical masses are inferred from the {\sc GalPak3D} modelling (Section~\ref{subsec:spectra+kinematics}) \label{tab:source_properties}}
\begin{center}
\begin{tabular}{@{}ll|cc @{}}
\hline \hline
Source & & ALESS 49.1 & ALESS 57.1 \\

\hline
\multicolumn{2}{l}{SED fitting} & \multicolumn{2}{c}{}\\
\hline
$L_{3-2000 \mathrm{\mu m}}$ & [$10^{12}$ L$_\odot$] & $7.1^{+0.8}_{-0.9}$ & $7.4^{+0.9}_{-0.9}$ \\
SFR & [M$_\odot$ yr$^{-1}$] & $490^{+30}_{-60}$ & $480^{+70}_{-60}$\\
$T_{dust}$ & [K] & $46^{+6}_{-2}$ & $51^{+7}_{-4}$ \\
$M_\star$ & [$10^{10}$ M$_\odot$] & $4.4^{+1.8}_{-0.3}$ & $4.3^{+1.7}_{-0.8}$ \\
$M_\mathrm{gas}\,^b$ & [$10^{10}$ M$_\odot$] & $5\pm2$ & $5\pm2$ \\
$M_\mathrm{dust}$ & [$10^8$ M$_\odot$] & $4.4^{+0.5}_{-0.7}$ & $4.1^{+0.5}_{-0.6}$ \\

\hline

\multicolumn{2}{l}{Line Luminosities} & \multicolumn{2}{c}{}\\
\hline
$L_\mathrm{[CII]}$ & [$10^9$ L$_\odot$] & 3.0$\pm$0.8 & 1.1$\pm$0.4 \\
$L'_\mathrm{[CII]}$ & [$10^{10}$ K km s$^{-1}$ pc$^2$] & 14$\pm$4 & 5.1$\pm$1.7 \\
$L_\mathrm{CO(3-2)}\,^a$ & [$10^9$ L$_\odot$] & 0.070$\pm$0.005 & 0.062$\pm$0.016 \\
$L'_\mathrm{CO(3-2)}\,^a$ & [$10^{11}$ K km s$^{-1}$ pc$^2$] & 0.51$\pm$0.04 & 0.05$\pm$0.01 \\
$L_\mathrm{CO(1-0)}\,^a$ & [$10^6$ L$_\odot$] & 2.1$\pm$0.2 & 2.6$\pm$0.7 \\
\hline

\multicolumn{2}{l}{Source sizes - single Gaussian} & \multicolumn{2}{c}{}\\
\hline
$R_{1/2}^{\mathrm{[CII]}}$ & [arcsec] & 0.163$\pm$0.013 & 0.101$\pm$0.010 \\
$S^{\mathrm{[CII]}}$ & [mJy] & 10.9$\pm$0.9 & 8.6$\pm$0.9 \\
$R_{1/2}^{160 \mu\mathrm{m}}$ & [arcsec] & 0.173$\pm$0.009 & 0.128$\pm$0.006 \\

$S^{160 \mu\mathrm{m}}$ & [mJy] & 11.2$\pm$0.4 & 7.6$\pm$0.4 \\
$R_{1/2}^\mathrm{CO}\,^a$ & [arcsec] & 0.33$\pm$0.5 & 0.39$\pm$0.06 \\

\hline
\multicolumn{2}{l}{Source sizes - two Gaussians} & \multicolumn{2}{c}{}\\
\hline
$R_{1/2}^{\mathrm{[CII]}}$ (compact) & [arcsec] & 0.128$\pm$0.015 & -- \\
$R_{1/2}^{\mathrm{[CII]}}$ (extended) & [arcsec] & 1.1$\pm$0.3 & -- \\
$S^{\mathrm{[CII]}}$ (compact) & [mJy] & 7.4$\pm$1.1 & -- \\
$S^{\mathrm{[CII]}}$ (extended) & [mJy] & 29$\pm$8 & -- \\
$R_{1/2}^{160 \mu\mathrm{m}}$ (compact) & [arcsec] & -- & 0.109$\pm$0.007 \\
$R_{1/2}^{160 \mu\mathrm{m}}$ (extended) & [arcsec] & -- & 0.67$\pm$0.11 \\

$S^{160 \mu\mathrm{m}}$ (compact) & [mJy] & -- & 5.8$\pm$0.5 \\
$S^{160 \mu\mathrm{m}}$ (extended) & [mJy] & -- & 7.6$\pm$1.4 \\

\hline

\multicolumn{2}{l}{[\ion{C}{2}] and CO(3--2) kinematics} & \multicolumn{2}{c}{}\\
\hline
$i_\mathrm{[CII]}$ & [deg] & 39$\pm$5 & 58$\pm$5 \\
$M_\mathrm{dyn}^\mathrm{[CII]}$ ($R\leq$2~kpc) & [$10^{10}$ M$_\odot$] & $6.2\pm5.5$ & $2.7\pm1.6$ \\
$M_\mathrm{dyn}^\mathrm{CO} (R<R_{1/2}^\mathrm{CO})$ & [$10^{10}$M$_\odot$] & $11 \pm 2$ & $11 \pm 5$ \\

\hline

\multicolumn{4}{l}{$^{a}$ \citet{CR18}: $L_{line}$ integrated over the entire line width}\\
\multicolumn{4}{l}{as opposed to integrating only over FWHM as in \citet{CR18}}\\
\multicolumn{4}{l}{$^{b}$ \citet{CR18}. $M_{gas}$ estimated assuming $\alpha_\mathrm{CO}$ = 1.0.}\\
\end{tabular}
\end{center}

\end{table*}

\newpage
\subsection{Spectral energy distribution modelling}
\label{subsec:SED_models}

We infer the global stellar and interstellar medium properties of ALESS 49.1 and 57.1 from the spatially-integrated spectral energy distributions (SEDs) using the {\sc MagPhys} package \citep{daCunha2008}, specifically its high-redshift extension \citep{daCunha2015}. These differ from the previously published SEDs \citep{daCunha2015} by using the CO(3--2)-derived spectroscopic redshifts compared to the photometric ones from \citet{daCunha2015}, and inclusion of ALMA Bands 3/4/8 continuum flux-densities. Namely, in addition to the Band~8 continuum measurements from Table~\ref{tab:targets+observed}, we include Band~4 continuum measurements ($S_\mathrm{2.1mm}$ = 380$\pm$100~$\mu$Jy and 65$\pm$50~$\mu$Jy for ALESS 49.1 and 57.1; da~Cunha et al., in prep.), as well as the ALMA Band~3 continuum measurement for ALESS~49.1, $S_\mathrm{3.3mm}=37\pm5$~$\mu$Jy \citep{Wardlow2018}. 

 The {\sc MagPhys}-inferred source properties are listed in Table~\ref{tab:source_properties}, the observed multi-wavelength photometry and {\sc MagPhys} spectral energy distribution models are provided in Appendix~\ref{sec:appendix_2}. The estimated dust temperatures $T_\mathrm{dust}=47^{+5}_{-9}$~K and $52^{+10}_{-6}$~K in ALESS~49.1 and 57.1 are warmer than the median $T_\mathrm{dust}$=42$\pm$2~K of the ALESS SMGs inferred from {\sc MagPhys} modelling \citep{daCunha2015}. The more precise spectroscopic redshifts and additional ALMA photometry result in a temperature increase compared to values reported by \citet{daCunha2015} $T_\mathrm{dust}=46^{+0}_{-3}$~K and $43^{+15}_{-14}$~K, respectively. Elevated $T_\mathrm{dust}$ in intensely star-forming $z\sim4.5$ galaxies was reported by \citet{Cooke2018}, who interpret the inferred median $T_\mathrm{dust}=55\pm4$~K as evidence for high $\Sigma_\mathrm{SFR}$ at high redshift. Note that the \citet{daCunha2015} and \citet{Cooke2018} models use \textit{Herschel} SPIRE and ALMA Band~7 photometry, while our models include ALMA Bands 3, 4 and 8 observations, thus better sampling the Rayleigh-Jeans tail of dust thermal spectrum. Compared to the {\sc MagPhys} models of the entire ALESS sample \citep{daCunha2015}, ALESS~49.1 and 57.1 have SFR higher by a factor of $\sim$2 (\citealt{daCunha2015}: median SFR = 280$\pm$70 M$_\odot$ yr$^{-1}$) and stellar mass factor of 2 lower (\citealt{daCunha2015}: $M_* = (8.9\pm0.1)\times10^{10}$~M$_\odot$ yr$^{-1}$); the dust mass is consistent with the median \citet{daCunha2015} value ($(5.6\pm1.0)\times10^{10}$~M$_\odot$). The gas depletion timescale $M_\mathrm{gas}$/SFR is 100$\pm$40~Myr in both ALESS~49.1 and 57.1, in line with $z\simeq2-4$ SMGs \citep{Huynh2017, Bothwell2013}, and a few times lower than claimed for $z=1-3$ massive main-sequence galaxies from the PHIBBS survey ($0.7\pm0.1$~Myr, \citealt{Tacconi2018}).

Using the SED models, we can estimate the fraction of the 160-$\mu$m continuum that is resolved out by comparing the observed rest-frame 160-$\mu$m continuum fluxes with SED modeling predictions, assuming constant $T_\mathrm{dust}$ and optical depth across the source. Namely, we use {\sc MagPhys} to perform SED modeling using all the photometry points apart from the rest-frame 160-$\mu$m continuum. The predicted 160-$\mu$m flux-densities are 11.8 and 7.6 mJy for ALESS~49.1 and 57.1, respectively; the observed flux densities match the predicted ones within $<$10\%. Therefore, we conclude that our observations recover the bulk of the 160-$\mu$m emission, in line with the compact continuum sizes inferred in Section~\ref{subsec:uv-plane}.

\section{Discussion}
\label{sec:discussion}


\subsection{Comparison with CO(3--2) emission}

Based on the $uv$-plane analysis in Section~\ref{subsec:uv-plane} which was tested on mock ALMA data in Section~\ref{subsec:mock_data}, we found the [\ion{C}{2}]/160-$\mu$m continuum surface brightness distribution to be dominated by a compact component, embedded within a low-surface-brightness, extended emission. We now compare these morphologies to the CO(3--2) surface brightness profiles, and other low- and high-redshift observations and simulations.

\subsubsection{160-$\mu$m continuum size}

Comparing the 160-$\mu$m continuum emission sizes with the CO(3--2) sizes (Table~\ref{tab:source_properties}), the 160-$\mu$m continuum is 1.9$\pm$0.3 (ALESS~49.1) and 3.1$\pm$0.7 (ALESS~57.1) more compact than the CO(3--2) emission.

Compact dust continuum emission embedded in a larger molecular gas reservoir has been observed in a number of high-resolution studies of $z=2-4$ SMGs. For example, PdBI/VLA imaging of GN20 ($z=4.05$) revealed a compact dusty, star-forming region within a large disc as traced by the CO(2--1) emission \citep{Hodge2012, Hodge2015}. At very high spatial resolutions ($\sim100$~pc), similar morphology has been seen in the strongly-lensed $z=3$ SMG SDP.81 \citep{alma2015}, which shows a compact ($\sim$3 kpc across), dusty star-forming disc \citep{Rybak2015a}, embedded in a large ($\geq10$~kpc) molecular gas reservoir, as traced by the CO(1--0) emission \citep{Valtchanov2011, Rybak2015b}. Similarly, \citet{Spilker2015} observed compact ALMA Band~7 continuum and extended CO(3--2)/(1--0) emission in two strongly lensed galaxies from the South Pole Telescope sample; note that in SMGs, the CO(3--2) is typically significantly less extended than the CO(1--0) line (e.g.,\,\citealt{Ivison2011,Riechers2011}).
Using ALMA 860-$\mu$m continuum, VLA CO(3--2) and SINFONI H$\alpha$ observations of ALESS~67.1 ($z$=2.1), \citet{Chen2017} found the ALMA continuum to be a factor of $\sim$5 more compact than CO~(3--2) and H$\alpha$ emission (which are similar in size). Finally, \citet{CR18} compared stacked CO(3--2) and \citet{Hodge2016} 860-$\mu$m ALMA observations of a sample of 4/18 SMGs from the ALESS sample, showing the FIR continuum to be more compact by a factor of $>$2. Based on a radiative transfer modelling, \citet{CR18} found that the compact FIR and extended CO(3--2) sizes are consistent with a decrease in dust temperature and optical depth towards the outskirts of the source. The compact continuum sizes in ALESS~49.1 and 57.1 thus add to the growing evidence for variations ISM conditions in SMGs on scales of few kpc.

\subsubsection{[CII] emission size}

Based on the source sizes inferred from $uv$-plane fitting, the [\ion{C}{2}] emission is $2.0\pm0.4$ (ALESS~49.1) and $3.9\pm0.7$ (ALESS~57.1) times more compact than CO(3--2). The compact size of the high surface-brightness [\ion{C}{2}] component contrasts with a relatively similar extent of the [\ion{C}{2}] and low-$J$ CO emission in local galaxies, as presented by \citet{deBlok2016}, who found the [\ion{C}{2}] to be ``\emph{slightly less compact than the CO}" (scale radius $\sim70\pm20$\% larger). Note that the \citet{deBlok2016} galaxies have $\Sigma_\mathrm{SFR}\simeq10^{-3}-10^{-2}$~M$_\odot$ yr$^{-1}$ kpc$^{-2}$, $>3$~dex lower than ALESS~49.1 and 57.1.

The [\ion{C}{2}] morphologies consisting of a high surface-brightness compact, and a low surface-brightness extended component have been observed in several local star-forming galaxies. For example, Kuiper Airborne Observatory imaging of the [\ion{C}{2}] emission in NGC~6946 \citep{Madden1993} resolved three distinct components: (1) a bright, compact nucleus ($R<300$~pc); (2) a faint, diffuse component at least 12~kpc in radius, contributing $\sim$40\% of the total [\ion{C}{2}] flux \citep{Contursi2002} and (3) local enhancements corresponding to the spiral arms.

Similarly, one of the best-studied $z\geq5$ sources with resolved [\ion{C}{2}]/low-$J$ CO observations -- a $z\sim$6.42 quasar SDSS J1148+5251 \citep{Walter2004, Walter2009} -- has a compact ($R\sim0.75$ kpc) [\ion{C}{2}] emission embedded within a much more extended ($R\sim2.5$ kpc) CO(3--2) reservoir \citep{Walter2009, Stefan2015}. Using sensitive PdBI observations, \citet{Cicone2015} found an extremely extended (out to $\sim$30 kpc) [\ion{C}{2}] emission associated with powerful outflows likely driven by the central engine in J1148+5251.

Apart from observational evidence, a compact, high surface-brightness [\ion{C}{2}] component in star-forming galaxies has been predicted by simulations. In particular, using zoom-in cosmological SPH simulations of several mildly star-forming $z\sim2$ galaxies (SFR=5--60~M$_\odot$ yr$^{-1}$), \citet{Olsen2015} predicted that the [\ion{C}{2}] emission is concentrated into a compact central region ($R\sim1$~kpc), with a much fainter (by up to 1~dex) extended component -- similar to our double-component model of [\ion{C}{2}] emission in ALESS~49.1. In the \citet{Olsen2015} simulations, the compact size of the [\ion{C}{2}] emission is due to gas inflows into the central star-forming regions, and the [\ion{C}{2}] emission is dominated by the dense, molecular phase of the ISM. This contrasts with low-SFR galaxies, where the bulk of the star-formation takes place in the spiral arms at large galactocentric radii (e.g.,\,\textit{Herschel} survey of the Milky Way [\ion{C}{2}] emission, \citealt{Pineda2013}).

\subsubsection{Inferring FIR source sizes using the Stacey et al.,~2010, relation}

Given the limited angular resolution of many high-redshift [\ion{C}{2}] detections, the source size can not be inferred directly from unresolved/marginally resolved data. However, parallel [\ion{C}{2}]/CO/FIR continuum observations have been used to constrain the source size, assuming the bulk of the line and continuum emission originates in PDR regions.

This technique, introduced by \citet{Stacey2010}, compares the observed [\ion{C}{2}]/CO/FIR fluxes with predictions from the PDR models of (\citealt{Kaufman1999}, see Section~\ref{sec:PDR_models}) to infer the FUV field strength $G_0$ and the density $n$. The observed $L_\mathrm{FIR}$ and inferred $G_0$ are then used to infer the FIR size $R_\mathrm{FIR}$ using the \citet{Wolfire1990} relations. In particular, \citet{Wolfire1990} distinguish two main regimes:

\begin{enumerate}

\item $\langle \lambda_\mathrm{FUV} \rangle < R_\mathrm{FIR}$

\begin{multline}
\langle G_0 \rangle= 3\times10^4 \Big(\frac{L_\mathrm{FIR}}{10^{10} \mathrm{L_\odot}}\Big) \Big(\frac{\langle \lambda_\mathrm{FUV} \rangle}{100 \mathrm{pc}}\Big)
 \Big(\frac{100 \mathrm{pc}}{R_\mathrm{FIR}}\Big)^3 \\(1-\exp(-R_\mathrm{FIR}/\langle \lambda_\mathrm{FUV} \rangle)),
\end{multline}

\item $\langle \lambda_\mathrm{FUV} \rangle \geq R_\mathrm{FIR}$

\begin{equation}
\langle G_0 \rangle= 3\times10^4 \Big(\frac{L_\mathrm{FIR}}{10^{10} \mathrm{L_\odot}}\Big)\Big(\frac{100 \mathrm{pc}}{R_\mathrm{FIR}}\Big)^2,
\end{equation}
\end{enumerate}

where $G_0$ is given in the units of the Habing field ($1.6\times10^{-3}$ erg cm$^{-2}$ s$^{-1}$) and $R_\mathrm{FIR}$ and the FUV photon mean free path $\langle \lambda_\mathrm{FUV} \rangle $ are in pc. \citet{Stacey2010} assume that $\langle \lambda_\mathrm{FUV} \rangle$ in $z=1-2$ star-forming galaxies is the same as in a nearby starburst M82. For M82, \citet{Stacey2010} assume $L_\mathrm{FIR}=2.8\times10^{10}$~L$_\odot$, $G_0=10^3$, $R_\mathrm{FIR}=150$~pc.

With high-resolution [\ion{C}{2}], CO(3--2) and FIR observations and robust source sizes in hand, we now investigate the applicability of the \citet{Stacey2010} relations to high-redshift SMGs. Assuming that $\langle \lambda_\mathrm{FUV} \rangle$ is smaller than $R_\mathrm{FIR}$ (an appropriate choice given the heavily obscured, dusty environment), we infer $R_\mathrm{FIR}=410\pm50$ and $420\pm70$~pc for ALESS~49.1 and 57.1, respectively; a factor of 2.5--3.5 smaller than the actual FIR half-light radii measured from the $uv$-plane fitting (Section~\ref{subsec:uv-plane}). We note that the $L_\mathrm{FIR}$, $G_0$ and $R_\mathrm{FIR}$ estimates for M82 from the literature show a considerable scatter; alternatively, the $\langle \lambda_\mathrm{FUV} \rangle$ in SMGs might be somewhat longer from that in the central region of M82. Crucially, if the FIR sizes of high-redshift SMGs are systematically underestimated by similar factors, the star-formation rate surface density $\Sigma_\mathrm{SFR}\propto L_\mathrm{FIR}/R_\mathrm{FIR}^2$ will be \emph{overestimated} by 0.5--1.0~dex - a shift that might affect a number of unresolved observations in e.g.,\,the [\ion{C}{2}]/FIR -- $\Sigma_\mathrm{SFR}$ plane (Figure~\ref{fig:CII_FIR_deficit}).

\subsection{Molecular gas kinematics}
\label{sec:kinematics_discussion}

\begin{figure*}
\begin{centering}
\includegraphics[height=6.0cm, clip=true]{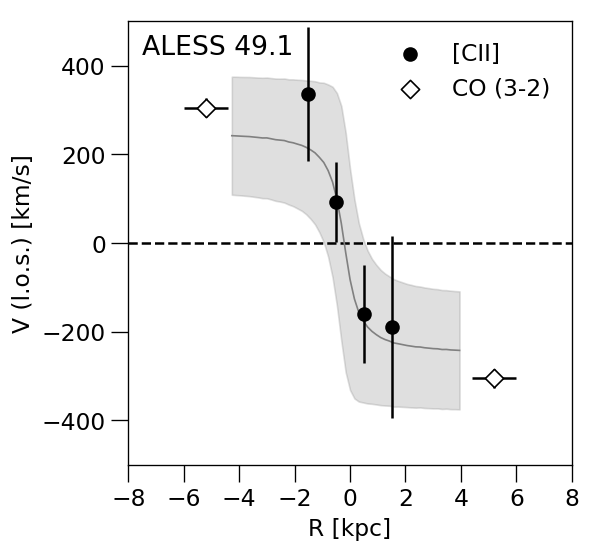}
\includegraphics[height=6.0cm, clip=true]{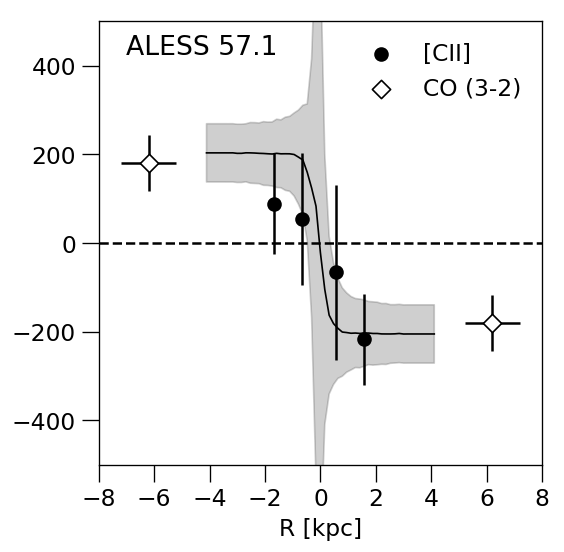}

\caption{Line-of-sight velocity curves for ALESS~49.1 and 57.1, showing the [\ion{C}{2}] velocity and velocity dispersion measured from the image cubes (full dots), with rotation curves inferred from {\sc GalPak3D} (full line, velocity dispersion estimate indicated by the shaded region), alongside the CO(3--2) measurements (empty diamonds, \citealt{CR18}). The [\ion{C}{2}] measurements were extracted along the major kinematic axis determined by the {\sc GalPak3D} modelling from an aperture one beamwidth wide. The line-of-sight velocity flattens out at $\sim$2-kpc radius, suggesting disc-dominated kinematics. The large {\sc GalPak3D}-predicted velocity dispersion at the centre of ALESS 57.1 is a numerical artefact. \label{fig:rot_curves} }
\end{centering}
\end{figure*}

The [\ion{C}{2}] and CO(3--2) lines provide two independent measurements of the molecular gas velocity structure. Due to their different spatial extents (Section~\ref{subsec:uv-plane}), the [\ion{C}{2}] and CO(3--2) emission trace the velocity field at different radii.
Figure~\ref{fig:rot_curves} shows the line-of-sight velocity as traced by the [\ion{C}{2}] and CO(3--2) emission, and the {\sc GalPak3D} model of the [\ion{C}{2}] disc. In particular, the [\ion{C}{2}] emission probes the velocity field within the inner 2-kpc region. Typically, rotational curves of disc-like galaxies are decomposed into contributions from the dark matter halo, and baryons in the form of a galactic bulge and disc. Given the baryonic mass $M_\mathrm{bar}=M_*+M_\mathrm{gas}$ (Table~\ref{tab:source_properties}) is comparable to the dynamical mass enclosed within twice the CO(3--2) half-light radius (5.2 and 6.0 kpc in ALESS~49.1 and 57.1, respectively; Table~\ref{tab:source_properties}), we conclude that the inner rotation curves are baryon-dominated.

In both ALESS~49.1 and 57.1, the line-of-sight velocity $v_\mathrm{los}$ flattens at 2--4~kpc radius. As a dominant bulge would cause the $v_\mathrm{los}$ to flatten rapidly on scales of few hundred pc (e.g.,\,\citealt{Sofue1999}) whereas the $v_\mathrm{los}$ profiles in ALESS~49.1 and 57.1 are still rising at $R\geq1$~kpc, we speculate that the inner ($R\leq2$~kpc) rotational curves in ALESS~49.1 and 57.1 do not yet have a significant bulge component and hence are disc-dominated. However, higher-SNR data are necessary to confirm this hypothesis.

The rotation curves have been studied at a comparable resolution in only a handful of SMGs. In this respect, ALESS~49.1 and 57.1 velocity fields are most directly comparable to those in $z=2.4$ twin hyperluminous SMGs H-ATLAS J084933 \citep{Ivison2013} and strongly lensed SMGs SMM J2135-0102 ($z=2.03$, \citealt{Swinbank2011}) and SDP.81 ($z=3.04$, \citealt{Dye2015,Rybak2015b}), which flatten out at $\sim$200~km s$^{-1}$ within the inner 2-kpc radius. The dynamical mass enclosed within the central 2-kpc radius region of ALESS~49.1 and 57.1 is $M_{dyn}$ ($R\leq$2~kpc) = $(6.2\pm5.5)\times10^{10}$ and $(2.7\pm1.6)\times10^{10}$ M$\odot$, respectively; although CO(3--2) observations provide mass estimates at 5--6~kpc radius, the large uncertainties prevent us from investigating the mass profiles of the two sources. The limited extent of the bright [\ion{C}{2}] component and the possibility that an extended [\ion{C}{2}] emission in ALESS~49.1 is resolved out highlight the difficulties of using very high-angular-resolution observations to trace the kinematics of the cold gas reservoir.

\subsection{The resolved [CII]-FIR ratio}
\label{sec:cii_deficit}

The high resolution of ALMA Band 8 data allows us to investigate the $L_\mathrm{[CII]}/L_\mathrm{FIR}$ ratio at $z\sim3$ on 1-kpc scales. Here, we focus on the \emph{central region} ($R\leq$2~kpc), in which the contribution of any extended [\ion{C}{2}] component is below 20\%. This discussion does not include the \emph{global} tracer ratios, as the Band~8 observations might resolve out significant fraction of the total [\ion{C}{2}] flux from an extended component.

The star-formation rate surface density $\Sigma_\mathrm{SFR}$ is estimated by adopting the global SFR from our SED fits \ref{subsec:SED_models}, assuming (1) a linear mapping between the rest-frame 160-$\mu$m emission and SFR; (2) no AGN contribution to the rest-frame FIR luminosity; (3) universal IMF (c.f \citealt{Zhang2018}). The linear mapping between the FIR continuum and SFR density assumes a constant dust temperature and opacity across the source. Further high-resolution, multi-band observations of the dust continuum and spatially-resolved SED modelling would be required for a more precise $\Sigma_\mathrm{SFR}$ estimate for different parts of the source. The [\ion{C}{2}]/FIR ratio was extracted by binning the [\ion{C}{2}] and 160-$\mu$m continuum maps (Fig.~\ref{fig:aless_images}) into pixels 1 beamwidth large; only pixels for which both the [\ion{C}{2}] and 160-$\mu$m continuum have SNR$\geq$2 are considered.

\begin{figure}
\begin{centering}
\includegraphics[width=8.5cm, clip=true]{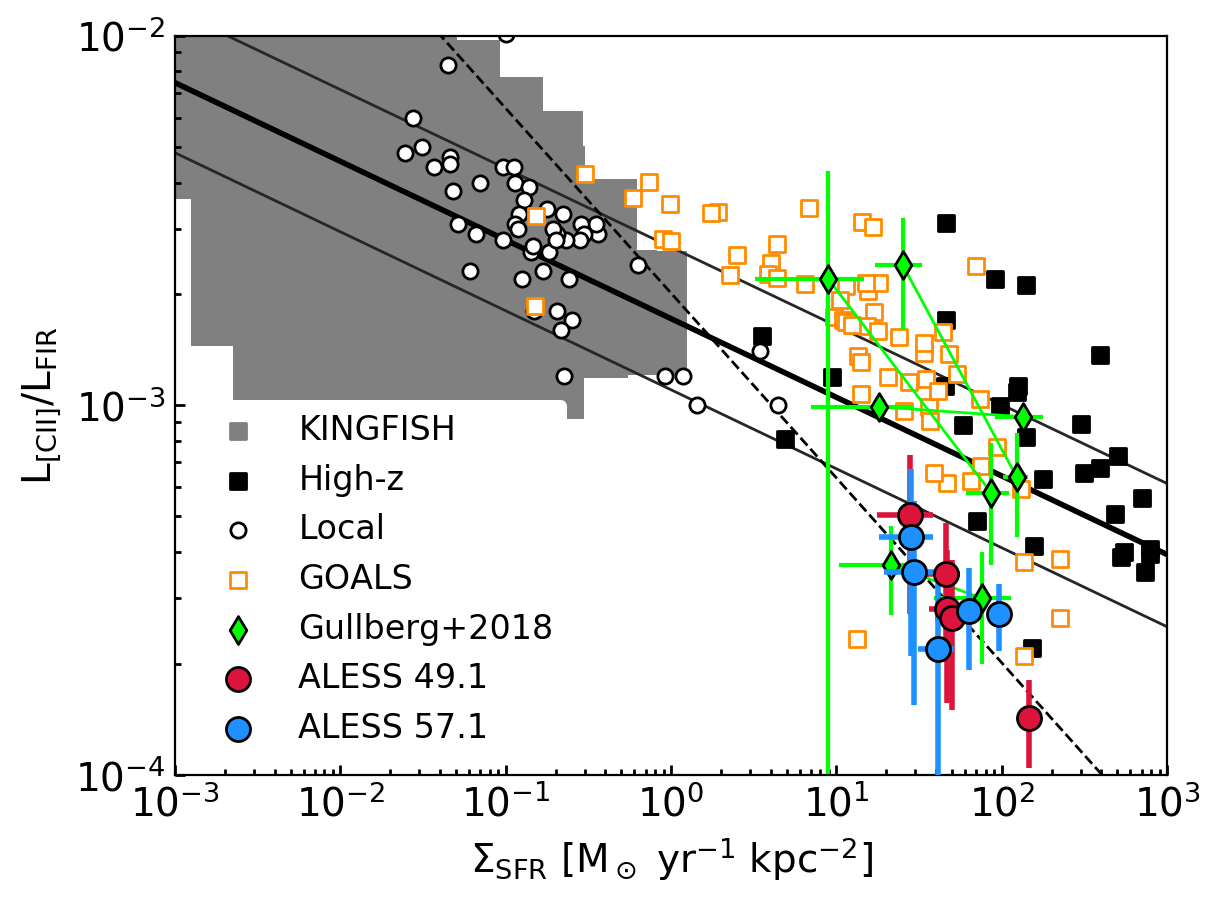}\\
\caption{Resolved [\ion{C}{2}]/FIR deficit in ALESS 49.1 and 57.1, compared to the {\sc Kingfish} \citet{Smith2017} and GOALS \citet{Diaz2017} samples and other local (empty circles) and high-redshift measurements (black squares), adapted from \citet{Smith2017} and \citet{Gullberg2018}. For each of the \citet{Gullberg2018} sources, we show the ratio for the inner and outer region, connected by a line. The thick black lines indicates the \citet{Smith2017} empirical fit to the [\ion{C}{2}]/FIR data (equation (\ref{eq:smith2017}) and the corresponding 1$\sigma$ scatter; the dashed line indicates the prediction from the \citet{MunozOh2016} temperature-saturation model (equation (\ref{eq:munozoh})), for $f_\mathrm{[CII]}$=0.13. The resolved ALESS~49.1 and 57.1 datapoints fall below the \citet{Smith2017} trend, and follow a much steeper slope, indicating a [\ion{C}{2}]-saturation regime. The errorbars on ALESS~49.1 and 57.1 measurements include a contribution from an extended [\ion{C}{2}]/160-$\mu$m continuum components. \label{fig:CII_FIR_deficit}}
\end{centering}
\end{figure}

Figure~\ref{fig:CII_FIR_deficit} shows the resolved $L_\mathrm{[CII]}/L_\mathrm{FIR}$ for ALESS 49.1 and 57.1 as a function of the SFR surface density $\Sigma_\mathrm{SFR}$, compared to resolved and unresolved measurements from both local and high-redshift sources (\citealt{Smith2017,Diaz2017,Gullberg2018} and references therein). In both ALESS~49.1 and 57.1, the resolved ratio decreases sharply with $\Sigma_\mathrm{SFR}$, confirming the [\ion{C}{2}]/FIR deficit on 1-kpc scales. 

Comparing our measurements with an empirical relation between [\ion{C}{2}]/FIR ratio and $\Sigma_\mathrm{SFR}$ proposed by \citet{Smith2017}, 

\begin{equation}
\Sigma_\mathrm{SFR} = (12.7\pm3.2)\Big(\frac{L_\mathrm{[CII]}/L_\mathrm{FIR}}{0.001}\Big)^{-4.7\pm0.8} \, \mathrm{M_\odot yr^{-1} kpc^{-2}}.
\label{eq:smith2017}
\end{equation}

we find that the bulk of the ALESS~49.1 and 57.1 [\ion{C}{2}]/FIR measurements are below the 1$\sigma$ scatter of the \citet{Smith2017} relation, with $\Sigma_\mathrm{SFR}$ almost 1--2 dex lower than those predicted by the \citet{Smith2017} relation. This suggests that the \citet{Smith2017} $\Sigma_\mathrm{SFR}$ -- $L_\mathrm{[CII]}/L_\mathrm{[CII]}$ might not be directly applicable in the high-$\Sigma_{SFR}$ regime in ALESS~49.1 and 57.1.

We note that the ALESS~49.1 and 57.1 [\ion{C}{2}]/FIR measurements fall below the redshift $\sim4.5$ resolved measurements of \citet{Gullberg2018}. This can be attributed to several factors: (1) different source selection; (2) aperture-averaging effects in \citet{Gullberg2018}, as the [\ion{C}{2}]/FIR ratio is calculated for apertures several kpc wide; (3) systematic uncertainties such as \citet{Gullberg2018} assuming $T_\mathrm{dust}=50\pm4$~K for all their sources. Regarding source selection, ALESS~49.1 and 57.1 have $L_\mathrm{FIR}$ more than 2$\times$ higher than \citet{Gullberg2018} sources, while being more compact in the rest-frame FIR continuum; consequently, our measurements might probe a higher $\Sigma_\mathrm{FIR}$ regime, which would correspond to lower [\ion{C}{2}]/FIR ratio and potentially higher $T_\mathrm{gas}$ (see Section~\ref{sec:PDR_models}). 

\begin{figure}
\begin{centering}
\includegraphics[width=8.5cm, clip=true]{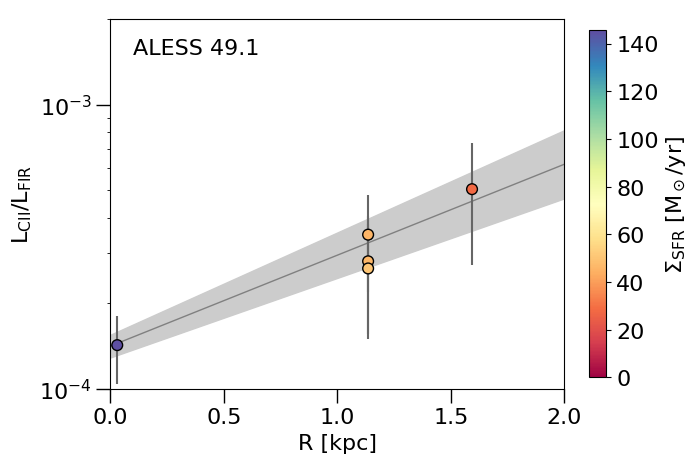}\\
\includegraphics[width=8.5cm, clip=true]{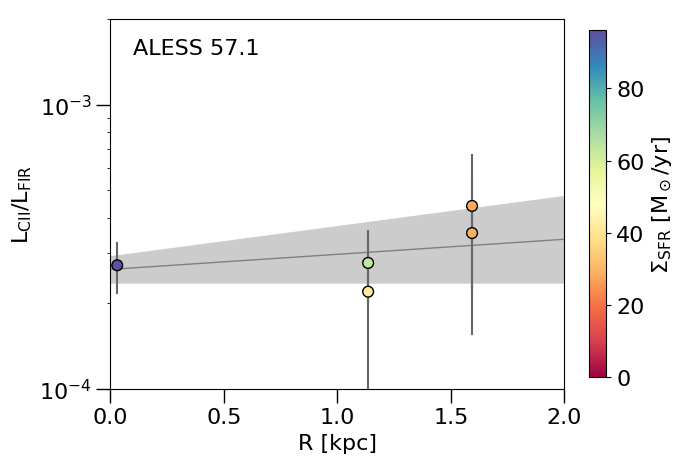}\\
\caption{[\ion{C}{2}]/FIR deficit radial trends in ALESS~49.1 and 57.1, colour-coded by the star-formation rate surface density $\Sigma_\mathrm{SFR}$; each point corresponds to a separate resolution element. The lines and shaded regions indicate the best-fitting slopes and corresponding uncertainties, respectively. The radial distances $R$ are measured from the 160-$\mu$m continuum surface brightness maximum. \label{fig:CII_FIR_radius}}
\end{centering}
\end{figure}

The radial variation of $L_\mathrm{[CII]}/L_\mathrm{FIR}$ in ALESS~49.1 and 57.1 is shown in Figure~\ref{fig:CII_FIR_radius}. Fitting the resolved [\ion{C}{2}]/FIR data with a power-law $L_\mathrm{[CII]}/L_\mathrm{[CII]}\propto R^\alpha$, we find a strong evidence for a radial variation of the [\ion{C}{2}]/FIR ratio in ALESS~49.1 ($\alpha = 0.41\pm0.06$), while in ALESS~57.1, the slope is consistent with being flat ($\alpha=0.05\pm0.05$). A decrease in the [\ion{C}{2}]/FIR ratio towards the centre of the source is seen in both local star-forming galaxies (e.g.,\,\citealt{Madden1993, Smith2017}) and high-redshift sources \citep{Gullberg2018} and indicates that the [\ion{C}{2}]/FIR deficit is driven by local processes, as opposed to global properties of the sources. In particular, \citet{Smith2017} found the [\ion{C}{2}]/FIR ratio to be suppressed by on average 30$\pm$15\% in the central $R\leq400$~pc regions of galaxies without an AGN. On the other hand, [\ion{C}{2}]/FIR ratio drops by a factor of a few over the inner 2~kpc in nuclear starbursts in M82 and M83 \citep{Herrera2018}. While the obscured \textit{Chandra}-detected AGN in ALESS~57.1 \citep{Wang2013} might be expected to suppress the [\ion{C}{2}] emission in the circumnuclear region, we do not detect any strong [\ion{C}{2}] suppression in ALESS~57.1 on 1-kpc scales. The difference in the [\ion{C}{2}]/FIR radial profiles in ALESS~49.1 and 57.1 is driven by the larger scatter in [\ion{C}{2}]/FIR ratio for a given $\Sigma_\mathrm{SFR}$ in ALESS~57.1 (see Figure~\ref{fig:CII_FIR_deficit}), which is a result of the complex [\ion{C}{2}] and 160-$\mu$m morphology in that source (Figure~\ref{fig:aless_images}). Note that the limited SNR of our data at $R\geq2$~kpc prevents us to study the [\ion{C}{2}]/FIR radial dependence at larger radii.

\subsection{Comparison with PDR models}
\label{sec:PDR_models}

The relative intensities of the [\ion{C}{2}], CO(3--2) and FIR emission from hot-dissociation regions (PDRs) depend on the ionizing FUV field strength $G_0$ and the gas density $n$(H), which determine the depth of the outer C$^+$ layer. We now use our resolved [\ion{C}{2}]/CO(3--2)/FIR observations to infer the $G_0$ and $n$ using the PDR models from the {\sc PDRToolbox} library \citep{Kaufman2006, Pound2008}. We focus on the central regions of the source ($R\leq2$~kpc) adopting the best-fitting models from Section~\ref{subsec:uv-plane}. 

For a proper comparison, several corrections need to be applied. First, the [\ion{C}{2}] emission from the ionized ISM needs to be subtracted from the observed [\ion{C}{2}] signal. The contribution from the ionized gas can be estimated from [\ion{N}{2}] 122/205-$\mu$m lines, which have similar critical density for collisions with electrons (300~cm$^{-3}$/32~cm$^{-3}$) as the [\ion{C}{2}] line (50~cm$^{-3}$, \citealt{Goldsmith2012}), but ionization energy $>13.6$~eV and hence traces only the ionized gas. \citet{Croxall2017} carried out a systematic study of the [\ion{N}{2}] 122/205~$\mu$m lines in a sample of 21 nearby star-forming galaxies, estimating the fraction of [\ion{C}{2}] in PDRs as $f_\mathrm{[CII]}^\mathrm{PDR}\geq0.8$ for sources with $\Sigma_\mathrm{SFR} \geq 10^{-2}$ M$_\odot$ yr$^{-1}$ kpc$^{-2}$. At high redshift, \citet{Zhang2018b} used [\ion{N}{2}] 122-$\mu$m line observations in $z=1-3$ SMGs to derive 10--15\% ionized gas contribution to [\ion{C}{2}] luminosity, assuming Galactic diffuse gas N and C abundancies. Consequently, we adopt a conservative (i.e.,\,low) estimate of $f_\mathrm{[CII]}^\mathrm{PDR} = 0.8$.

Second, {\sc PDRToolbox} models are derived for a one-dimensional, semi-infinite slab, illuminated from the face side only. In the intensely star-forming SMGs, we expect the clouds to be illuminated from multiple sides and the optical thickness of individual tracers needs to be considered. Namely, for optically thick emission, only the emission from the side of the cloud facing the observer is observed; for the optically thin emission, the distant side of the cloud adds to the observed fluxes and the line intensities expected from PDR models need to be multiplied by a factor of 2 \citep{Kaufman1999}. The 160-$\mu$m continuum is generally assumed to be optically thin. Based on the models of \citet{Narayanan2014}, CO(3--2) line is expected to be optically thick with a median optical depth $\geq10$ for $\Sigma_\mathrm{SFR}=10^{-1}-10^{3}$~M$_\odot$ yr$^{-1}$. The [\ion{C}{2}] emission is optically thin in most environments; although there is evidence for moderately optically thick [\ion{C}{2}] (optical depth $\sim$1) in both Galactic PDRs \citep{Graf2012, Sandell2015} and high-redshift sources \citep{Gullberg2015}. Therefore, for an optically thin [\ion{C}{2}], the predicted [\ion{C}{2}]/CO(3--2) ratio has to be increased 2$\times$; for an optically thick [\ion{C}{2}], the predicted [\ion{C}{2}]/FIR ratio has to be reduced by 1/2. We consider both optically-thin and optically-thick [\ion{C}{2}] scenarios; however, as the evidence for optically-thick [\ion{C}{2}] is limited, we adopt the values derived for the optically-thin ]\ion{C}{2}] for the rest of this paper.

\begin{table}
\caption{Comparison with {\sc PDRToolbox} PDR models: observed [\ion{C}{2}]/CO(3--2) and [\ion{C}{2}]/FIR continuum ratios, along with inferred FUV field strength $G_0$, density $n$(H) and PDR surface temperature $T_\mathrm{PDR}$ for ALESS~49.1 and 57.1. We consider both optically thin and thick [\ion{C}{2}] scenarios. The [\ion{C}{2}] luminosities are given before subtracting the contribution from non-PDR sources. \label{tab:pdr_models}}

\begin{center}
 \begin{tabular}{@{}llcc@{}}
 \hline \hline
Source & & ALESS 49.1 & ALESS 57.1 \\
\hline
{[\ion{C}{2}]}/FIR & & $(3.3\pm1.5)\times10^{-4}$ & $(3.4\pm2.8)\times10^{-4}$ \\
{[\ion{C}{2}]}/CO(3--2) & & $310\pm100$ & $600\pm240$ \\
\hline
\multicolumn{4}{l}{[\ion{C}{2}] optically thin}\\
\hline
$\log G_0$ & & $4.1^{+0.3}_{-0.3}$ & $4.2^{+0.9}_{-0.4}$ \\
$\log n$(H) & [cm$^{-3}]$ & $4.7^{+0.4}_{-0.2}$ & $4.1^{+0.5}_{-0.4}$ \\
$T_\mathrm{PDR}$ & [K] & $720\pm200$ & $670\pm170$ \\
\hline
\multicolumn{4}{l}{[\ion{C}{2}] optically thick}\\
\hline
$\log G_0$ & & $3.8^{+0.3}_{-0.2}$ & $3.7^{+1.0}_{-0.3}$ \\
$\log n$(H) & [cm$^{-3}]$ & $3.9^{+0.4}_{-0.2}$ & $3.4^{+0.4}_{-0.3}$ \\
$T_\mathrm{PDR}$ & [K] & $430\pm30$ & $480\pm90$ \\
\hline

\end{tabular}
\end{center}
\end{table}

\begin{figure}
\begin{centering}
\includegraphics[width=8.5cm, clip=true]{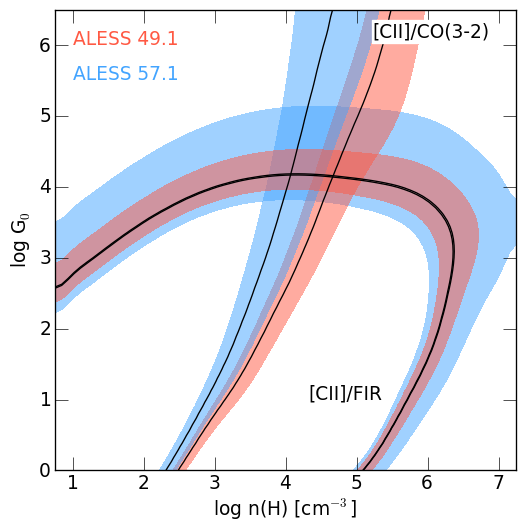}

\caption{Constraints on the FUV field strength $G_0$ and density $n$ for the central $R\leq2$~kpc region derived from comparison with \citet{Kaufman2006} PDR models. For the [\ion{C}{2}]/FIR/CO(3--2) emission, we adopt surface brightness profiles derived from the $uv$-plane fitting. The [\ion{C}{2}] emission is assumed to be optically thin, with PDR contributing 80\% of the observed signal. The coloured areas indicate 1$\sigma$ confidence regions. For the central regions of ALESS~49.1 and 57.1, we infer $G_0\sim10^4$ and $n=10^{4}-10^{5}$ cm$^{-3}$. \label{fig:pdr_models}}
\end{centering}
\end{figure}

Figure~\ref{fig:pdr_models} shows the contours in the $G_0$/$n$ space for the central $R\leq$2~kpc region of ALESS~49.1 and 57.1, with best-fitting values listed in Table~\ref{tab:pdr_models}. The combination of the three tracers provides orthogonal constraints on $G_0$ and $n$. In particular, $G_0$ is largely determined by the [\ion{C}{2}]/FIR ratio, and $n$ by [\ion{C}{2}]/CO~(3--2). Table \ref{tab:pdr_models} lists the inferred $G_0$ and $n$ values for the optically thin and optically thick [\ion{C}{2}] scenarios, along with the PDR surface temperature. For the optically-thin [\ion{C}{2}] case, the conditions in ALESS~49.1 and 57.1 are almost identical, with $G_0 \sim 10^4$ and $n \sim 10^4-10^5$~cm$^{-3}$, implying PDR surface temperature $T_\mathrm{PDR}$ of $\sim$700~K. Increasing the fraction of [\ion{C}{2}] emission from the PDRs from 0.8 to 1.0 causes the inferred $G_0$ and $n$ values to decrease by $\leq$0.25 dex. For the optically-thick [\ion{C}{2}] case, the inferred $G_0$ and $n$ decrease by up to 0.5 and 1.0~dex, respectively; $T_\mathrm{PDR}$ is reduced to 400--500~K. 

The $G_0$, $n$ values in the central region of ALESS~49.1 and 57.1 are comparable to $G_0 = 10^3 - 10^{4.5}$, $n=10^2-10^4$~cm$^{-3}$ inferred from unresolved observations of larger SMG samples, such as [\ion{C}{2}]/CO study of \citet{Stacey2010}, [\ion{C}{1}]/CO study of 14 $z\geq2$ SMGs by \citet{AZ2013} and FIR spectroscopy of lensed SMGs \citep{Wardlow2017, Zhang2018b}). Using [\ion{C}{2}] and low-$J$ CO observations in a sample of strongly lensed SMGs, \citet{Gullberg2015} found a larger scatter of FUV strength ($G_0 = 10^2 - 8\times10^3$) and density ($n=10^2-10^5$~cm$^{-3}$), although the effect of differential magnification might be substantial. For nearby star-forming galaxies, a comparison of observed spatially integrated [\ion{C}{2}], [\ion{O}{1}] 63-$\mu$m and FIR luminosities with the {\sc PDRToolbox} models was carried out by \citet{Malhotra2001} and \citet{Diaz2017}. ALESS~49.1 and 57.1 are consistent with the high-density \citet{Malhotra2001} sources; however, $G_0$ and $n$ in ALESS~49.1 and 57.1 are higher than in the most dense ultra-luminous infrared galaxies (ULIRGs) from the \citet{Diaz2017} sample, which have $G_0\sim10^3$, $n=1-10^3$~cm$^{-3}$. Note that the globally-averaged $G_0$ and $n$ in ALESS~49.1 and 57.1 might be lower than those inferred from the $R\leq2$~kpc region. Finally, compared to the resolved kpc-scale observations of local starburst galaxies NGC~6946 and NGC~1313 with inferred $G_0 = 10^3 - 10^4$, $n=10^{3.0}-10^{3.5}$~cm$^{-3}$ \citep{Contursi2002}, the central regions of ALESS~49.1 and 57.1 show similar $G_0$ and somewhat higher $n$.

What drives the strong FUV fields in ALESS~49.1 and 57.1: a central AGN, or star-formation? Although \textit{Chandra} X-ray observations of ALESS~49.1 and 57.1 \citep{Wang2013} revealed an obscured AGN in ALESS~57.1 (no emission from ALESS~49.1 was detected), it is unlikely that an obscured AGN would be driving a strong FUV field on few-kpc scales. On the other hand, the $G_0$ in the vicinity of \ion{H}{2} regions is of the order $10^3-10^5$ (e.g.,\,\citealt{Tielens1985,HollenbachTielens1999}), comparable to the values inferred from our PDR models. Similarly, typical $G_0$ and $n$ values for Galactic star-forming regions are of the order of $G_0 = 10^3 - 10^5$, $n=10^3-10^6$ cm$^{-3} $\citep{Stacey1991, Stacey2010}. We therefore conclude that the strong FUV field in ALESS~49.1 and 57.1 is due to star formation, rather than a central AGN.

\subsection{Origin of the [CII]/FIR deficit}

Having estimated $G_0$ and $n$ in the central regions of ALESS~49.1 and 57.1, we now turn to the mechanism driving the [\ion{C}{2}]/FIR deficit. We focus on the thermal saturation model proposed by \citet{MunozOh2016}, and the reduction of the photoelectric heating of the gas by small dust grains (e.g.,\,\citealt{BakesTielens1994,Malhotra2001}); we briefly discuss other potentially relevant mechanisms -- AGN contribution and dust-bounded HII regions -- in Section 4.5.3. For a more exhaustive list of proposed mechanisms for the [\ion{C}{2}]/FIR deficit, we refer the reader to \citet{Smith2017}.

\subsubsection{Thermal saturation of the [CII] line}

\citet{MunozOh2016} have proposed the thermal saturation of the upper level of the C$^+$ fine-structure transition as the main driver of the [\ion{C}{2}]/FIR deficit. In other words, when $T_\mathrm{gas}$ exceeds the C$^+$ ionization temperature (92~K), the upper/lower level population ratio (and the [\ion{C}{2}] luminosity) depends only weakly on $T_\mathrm{gas}$, while the FIR luminosity keeps on increasing. 

Our {\sc PDRToolbox} models imply high FUV fields strength ($G_0\sim10^4$) and densities ($n=10^{4}-10^5$~cm$^{-3}$) with gas surface temperatures larger than 500~K (Table \ref{tab:pdr_models}. The [\ion{C}{2}] transition is saturated in this regime. Following \citet{MunozOh2016}, the thermal cooling rate per Hydrogen atom via the [\ion{C}{2}] line depends on $T_\mathrm{gas}$ via

\begin{equation}
\Lambda_\mathrm{[CII]} = 2.3\times10^{-6} \, k_\mathrm{B} \, T_\mathrm{[CII]} \frac{2}{2+\exp(T_\mathrm{CII}/T_\mathrm{gas}) \frac{C}{H}},
\label{eq:cii_cooling}
\end{equation}

where $k_\mathrm{B}$ is the Boltzmann constant, $T_\mathrm{[CII]}$ is the [\ion{C}{2}] ionization temperature and C/H the relative abundances of the Carbon and Hydrogen atoms. Following the equation (\ref{eq:cii_cooling}), the [\ion{C}{2}] cooling rate increases by only $\sim$40\% between $T_\mathrm{gas}$=100~K and 500~K, whereas the $L_\mathrm{FIR}$ increases by a factor of few hundred, assuming $T_\mathrm{dust}$ scales proportionally with $T_\mathrm{gas}$ and $L_\mathrm{FIR}\propto T_\mathrm{dust}^{4+\beta}$, where the dust opacity $\beta$ is typically assumed to range from 1.5 to 2.5 (e.g.,\,\citealt{Casey2014}).

For a more direct comparison with \citet{MunozOh2016} model, we compare the resolved [\ion{C}{2}]/FIR observations in ALESS~49.1 and 57.1 (Figure~\ref{fig:CII_FIR_deficit}) to the predicted $\Sigma_\mathrm{SFR}$-[\ion{C}{2}]/FIR slope. According to \citet{MunozOh2016}, $L_\mathrm{[CII]}/L_\mathrm{FIR}$ ratio scales as:

\begin{equation}
\frac{L_\mathrm{[CII]}}{L_\mathrm{FIR}} = 2.2\times10^{-3} \frac{f_\mathrm{CII}}{0.13} \Big(\frac{\Sigma_\mathrm{FIR}}{10^{11} \mathrm{L}_\odot \mathrm{kpc}^{-2}} \Big)^{-1/2},
\label{eq:munozoh}
\end{equation}

where $f_\mathrm{CII}$ is the fraction of gas emitting in [\ion{C}{2}]. Fitting our datapoints with a power-law following the equation (\ref{eq:munozoh}), we obtain a best-fitting slope of $\gamma = -0.53\pm0.12$. This is in agreement with the thermal-saturation model slope of $\gamma = -0.5$ (equation (\ref{eq:munozoh})). 

We note that \citet{Diaz2017} discount the thermal saturation of the [\ion{C}{2}] line as a source of the [\ion{C}{2}]/FIR deficit in local star-forming galaxies. Namely, comparing the [\ion{O}{1}] 63~$\mu$m and [\ion{C}{2}] line ratios with a statistical equilibrium radiative transfer model, they obtain a scaling between dust and gas kinetic temperature $T_\mathrm{gas} = 1.6 - 2.1 \times T_\mathrm{dust}$. Given $T_\mathrm{dust}$ of 21--48~K, they find $T_\mathrm{gas}\leq$92~K, i.e. below the thermal-saturation regime. However, ALESS~49.1 and 57.1 show relatively high global $T_\mathrm{dust}$ = $47^{+5}_{-9}$ and $52^{+10}_{-6}$~K, respectively (Table~\ref{tab:source_properties}). Given the conversion factors from \citet{Diaz2017} and evidence for increase in dust and gas temperature towards the centre of SMGs \citep{CR18}, we conclude that $T_\mathrm{gas}$ in the central regions of ALESS~49.1 and 57.1 likely exceeds 92~K.

\subsubsection{Suppression of the [CII] emission due to positive grain charging}

In addition to the thermal saturation, another potentially important effect at the $G_0$, $n$ values inferred from our PDR analysis is the reduced photoelectric heating of the gas by electrons ejected from the small dust grains by the FUV photons (e.g.,\,\citealt{BakesTielens1994, Malhotra2001,Wolfire2003}). At high $G_0/n$ ratios, the grains become positively charged, thus raising the potential barrier for the electrons to escape. 

Qualitatively, a reduced photoelectric heating will manifest in moderate $T_\mathrm{gas}/T_\mathrm{dust}$ ratios. Although ALESS~49.1 and 57.1 have elevated dust temperatures compared to other high-redshift SMGs \citep{Swinbank2014,daCunha2015} and local ULIRGs \citep{Diaz2017}, the high PDR surface temperatures indicate that the gas is already heated to high temperature, at which point the [\ion{C}{2}] line becomes saturated.

Quantitatively, following \citet{Wolfire2003}, the photoelectric heating rate per hydrogen atom is give as:

\begin{multline}
\Gamma_\mathrm{PE} = \frac{1.1\times10^{-25}G'Z_d}{1+3.2\times10^{-2}\Big[G'\big(T_\mathrm{gas}/100~K\big)^{1/2} n_e^{-1}\phi_\mathrm{PAH}\Big]^{0.73}} \\ \mathrm{erg \, s^{-1}},
\label{eq:photoel}
\end{multline}

where $G'=1.7 \times G_0$, $Z_d$ is the dust-to-gas ratio (normalized to the Galactic value), $n_e$ the electron density and $\phi_\mathrm{PAH}$ a factor associated with the PAH molecules. The second term in the denominator corresponds to the positive grain charging. We follow \citet{MunozOh2016} by adopting $n_e = 1.1\times10^{-4} n$, $\phi_{PAH}=0.5$. Note that equation (\ref{eq:photoel}) assumes $T_\mathrm{gas}\leq$1000~K, which is satisfied for both ALESS~49.1 and 57.1. While for the ALESS~49.1 and 57.1 values of $G_0$, $n$ the second term in the denominator corresponding to the reduction in photoelectric heating, becomes dominant, the numerator is also proportional to $G_0$. Comparing a typical nearby star-forming galaxy with $G_0=10^2$, $n=10^4$~cm$^{-3}$, (c.f.\,\citealt{MunozOh2016}) and ALESS 49.1 and 57.1 with $G_0=10^4$, $n=10^4$~cm$^{-3}$, the second term in the determinant increases from $\sim$0.3 to $\sim18$, indicating a significant reduction in the gas heating due to grain charging. However, at the same time, the overall $\Gamma_\mathrm{PE}$ \emph{increases} by a factor of $\sim$6.

Therefore, we attribute the pronounced [\ion{C}{2}]/FIR deficit in the central regions of ALESS~49.1 and 57.1 to the high gas temperature which results in a quantum-level saturation of the C$^+$ fine structure.

\subsubsection{Other mechanisms for [CII]/FIR deficit}

Finally, we briefly consider other proposed mechanisms for the [\ion{C}{2}] deficit.

AGNs can contribute to the [\ion{C}{2}] deficit, both by increasing the FIR luminosity and reduce the C$^+$ abundance and the [\ion{C}{2}] emission by ionizing the carbon atoms to higher ionization states (C$^{2+}$, C$^{3+}$ etc.) via soft X-ray radiation \citep{Langer2015}. \textit{Chandra} observations of ALESS~49.1 and 57.1 revealed an obscured AGN with an extinction-corrected X-ray luminosity of $\log L_\mathrm{0.5-8.0\,keV} = 44.3$ and no X-ray emission in ALESS~49.1 \citep{Wang2013}. According to \citet{Langer2015} models, for $n\sim10^3$~cm$^{-3}$ (model closest to the conditions in ALESS~49.1 and 57.1), a 10\% decrease in the fraction of carbon in the C$^+$ state requires an X-ray flux of $f_X \simeq 10^{3.5}$~erg cm$^{-2}$ s$^{-1}$. Assuming an X-ray flux dilutes with distance $D$ as $1/4\pi D^2$, the AGN in ALESS~57.1 will affect only the innermost $\sim$100~pc radius, well below the spatial resolution of our data. Consequently, we do not expect a significant AGN contribution to the observed [\ion{C}{2}] deficit.

Another proposed explanation for the [\ion{C}{2}]/FIR deficit is the increased absorption of ionizing UV photons by the dust in dust-bounded \ion{H}{2} regions, which would result in increased FIR and decreased [\ion{C}{2}] luminosity, respectively \citep{Luhman2003}. \citet{Abel2009} used radiative transfer models of dust-bounded \ion{H}{2} region to qualitatively reproduce the [\ion{C}{2}]/FIR deficit trend. However, for the observed [\ion{C}{2}]/FIR values in ALESS~49.1 and 57.1, the \citet{Abel2009} models require densities of $n\leq1-2\times10^3$~cm$^{-3}$, i.e.\, much lower than those inferred from the PDR modelling. Furthermore, as already noted by \citet{MunozOh2016}, the dust drift time for high $G_0$, $n$ values becomes very short compared to O/B stars lifetime. We follow \citet{Draine2011} to estimate a dust drift time for a cluster of $10^3$ O/B stars, providing an ionizing photons flux $Q_0=10^{52}$~s$^{-1}$. Given the density of $10^{3.5}-10^{4.0}$~cm$^{-3}$, the dust drift time becomes $t_\mathrm{drift}=1.0-1.5\times10^5$~yr (Figure~9 of \citealt{Draine2011}). Even if the \ion{H}{2} regions in ALESS~49.1 and 57.1 are originally dust-bounded, given the long duration of the starburst compared to $t_\mathrm{drift}$, we do not expect a significant fraction of them to be dust-bounded at a given moment and hence do not expect the dust-bounded \ion{H}{2} regions to dominate the [\ion{C}{2}]/FIR deficit in ALESS~49.1 and 57.1.

\section{Conclusions}
\label{sec:conclusions}

We have investigated the morphology and kinematics of the [\ion{C}{2}] 157.74-$\mu$m line emission and associated 160-$\mu$m rest-frame continuum in two $z\sim3$ sources from the ALESS sample, based on the 0.15~arcsec ALMA Band~8 imaging. The morphology and [\ion{C}{2}] velocity field in both galaxies is consistent with an inclined rotating exponential disc. The [\ion{C}{2}] rotation curves show a flattening within the inner 2-3 kpc radius, indicative of a potential dominated by a baryonic disc.

Comparing the resolved maps of the [\ion{C}{2}] emission with those of CO(3--2) \citep{CR18}, we found the [\ion{C}{2}] surface brightness to be concentrated into a region a factor of 2--3 more compact the CO(3--2). In ALESS~49.1, we found evidence for a low surface-brightness, extended ($R_{1/2}\sim8$~kpc) [\ion{C}{2}] component, accounting for up to 80~percent of the [\ion{C}{2}] brightness. Based on mock ALMA observations, we excluded the possibility that [\ion{C}{2}] and 160-$\mu$m continuum follow the same single-Gaussian surface brightness as the CO(3--2) emission.

We compared of the [\ion{C}{2}]/FIR and CO~(3--2) observations to the {\sc PDRToolbox} photo-dissociation regions models \citep{Kaufman2006,Pound2008}. These indicate intense FUV radiation field ($G_0 \sim 10^4$) and moderately high gas densities ($n$(H) $=10^4-10^5$ cm$^{-3}$), comparable to the $G_0$ and $n$ values found in the central regions of nearby starbursts (e.g.,\,\citealt{Contursi2002}), as well as in other $z>2$ SMGs \citep{Stacey2010}. We attribute the strong FUV field to massive the star-formation, rather than an obscured AGN.

We tested the applicability of the \citet{Stacey2010} technique for estimating FIR source size from unresolved [\ion{C}{2}]/low-$J$ CO / FIR observations to ALESS~49.1 and 57.1. The \citet{Stacey2010} method yields FIR sizes factor of 2.5--3.5 more compact than measured from the $uv$-plane fitting; this bias causes the SFR surface-density to be overestimated by up to 1~dex, having a potentially significant impact on the interpretation of low-resolution observations.

Both ALESS~49.1 and 57.1 show a pronounced [\ion{C}{2}]/FIR deficit, with $L_\mathrm{[CII]}/L_\mathrm{FIR}=10^{-4}-10^{-3}$. The resolved [\ion{C}{2}]/FIR luminosity ratios fall below the empirical trend of \citet{Smith2017}, indicating a change in physical conditions compared to the nearby star-forming galaxies. A comparison with PDR models indicated surface temperatures of 400--800~K; at such a high temperature, the occupancy of the upper fine-structure level of C$^+$ ions (and the [\ion{C}{2}] luminosity) saturates, while FIR luminosity increases sharply. The most direct interpretation is that the strong [\ion{C}{2}] deficit is a result of the C$^+$ fines-structure thermal saturation \citep{MunozOh2016}. In addition, the resolved [\ion{C}{2}]/FIR measurements in ALESS~49.1 and 57.1 scale with star-formation rate surface density as $\Sigma_\mathrm{SFR}^{-0.53\pm0.12}$, in agreement with the thermal-saturation scenario slope of -0.5 \cite[]{MunozOh2016}. Although the photoelectric heating of the gas is reduced due to positive grain charging, for the $G_0$, $n$ values in ALESS~49.1 and 57.1, the thermal saturation is the main driving mechanism of the [\ion{C}{2}]/FIR deficit. This contrasts with the local star-forming galaxies, which are found to have gas temperatures below the C$^+$ ionization energy (e.g.,\,\citealt{Diaz2017}).

With only two galaxies in our sample, it is difficult to generalize our conclusions to the entire population of submillimeter galaxies. With ALMA now enabling routine observations of [\ion{C}{2}] emission at redshift 3 and beyond, and with a rapid increase of high-redshift sources with robust spectroscopic redshifts that are necessary for parallel [\ion{C}{2}]/CO observations, this study is a precursor to future multi-tracer, resolved studies of ISM at high redshift, and a necessary stepping stone to interpreting the [\ion{C}{2}] observations at very high redshift.

\section*{Acknowledgements}

The authors thank Frank Israel and Desika Narayanan for helpful discussions about the [\ion{C}{2}]/CO extent.

This paper makes use of the following ALMA data: ADS/JAO.ALMA\#2013.1.00470.S, \#2015.1.00019.S, \#2015.1.00948.S and \#2016.1.00754.S. ALMA is a partnership of ESO (representing its member states), NSF (USA) and NINS (Japan), together with NRC (Canada), MOST and ASIAA (Taiwan), and KASI (Republic of Korea), in cooperation with the Republic of Chile. The Joint ALMA Observatory is operated by ESO, AUI/NRAO and NAOJ. MR and JH acknowledge support of the VIDI research programme with project number 639.042.611, which is (partly) financed by the Netherlands Organisation for Scientific Research (NWO). IRS acknowledges support from the ERC Advanced Grant DUSTYGAL (321334) and STFC (ST/P000541/1). HD acknowledges financial support from the Spanish Ministry of Economy and Competitiveness (MINECO) under the 2014 Ram\'on y Cajal program MINECO RYC-2014-15686. JLW acknowledges support from an STFC Ernest Rutherford Fellowship (ST/P004784/1). 

\newpage

\appendix
\section{Companion sources in LESS~49 field}
\label{sec:appendix_1}

\citet{Hodge2013} identified a nearby counterpart to ALESS~49.1 - ALESS 49.2 (J2000 03:31:24.47 -27$^{\circ}$ 50' 38.1''). Detected at 4$\sigma$ confidence level in 870~$\mu$m continuum ($S_\mathrm{870um}=1.80\pm0.46$~mJy), it is included in the ``main'' ALESS sample. Additionally, ALESS~49.2 was detected in 3.3-mm continuum in ALMA Band~3 observations of \citet{Wardlow2018} with $S_\mathrm{3.3mm}=28\pm6$~$\mu$Jy; however, \citet{Wardlow2018} do not detect any CO(3--2) emission from ALESS~49.2, suggesting it is offset in redshift from ALESS~49.1. Finally, 1.4~GHz VLA observations (\citealt{Biggs2011}, based on \citealt{Miller2008} data, $\sim$2~arcsec resolution) detect radio continuum emission from ALESS~49.2 at $\sim4.5\sigma$ significance. The $\geq4\sigma$ detections in these high-resolution observations confirm that ALESS~49.2 is a physical source, rather than an imaging artifact.

We do not find any significant Band~8 continuum or [\ion{C}{2}] emission within a 1-arcsec radius of the position reported by \citet{Hodge2013}. Given the small size of the ALMA Band~8 primary beam (FWHM=14.1~arcsec) and the large distance of ALESS~49.2 from the phase tracking centre ($\sim$9.6~arcsec), the emission from ALESS~49.2 will be attenuated by $\sim$70\%. Therefore, we impose a 3$\sigma$ upper limit of 1.2~mJy on the ALESS~49.2 620-$\mu$m continuum flux-density. For a $z=3$ source, this constraint is compatible with a modified black-body SED with $T_\mathrm{dust}\simeq$20~K.

In addition to ALESS~49.2, \citet{Wardlow2018} detected significant 3.3-mm continuum emission from two additional sources in the vicinity of ALESS~49.1 - ALESS 49.L and ALESS49.C. However, we do not detect any emission at $\geq4\sigma$ significance in either Band~8 continuum or [\ion{C}{2}] emission at the position of any of the \citet{Wardlow2018} sources. Accounting for the primary beam response, we put 3$\sigma$ upper limits of $S_\mathrm{620um}\leq$ 0.4~mJy, for both ALESS~49.L and ALESS 49.C. Given that ALESS 49.L and ALESS 49.C are detected at $\sim8\sigma$ detection in the CO(3--2) and 3.3-mm continuum, respectively \citep{Wardlow2018}, we consider both sources to be physical. The companion sources in the LESS~49 field will be addressed in more detail using ALMA Band~4 observations in da Cunha et al. (in prep.).

\newpage
\section{Spectral energy distribution for ALESS~49.1 and 57.1}
\label{sec:appendix_2}

\begin{figure*}[h]
\begin{centering}
\includegraphics[width=14cm, clip=true]{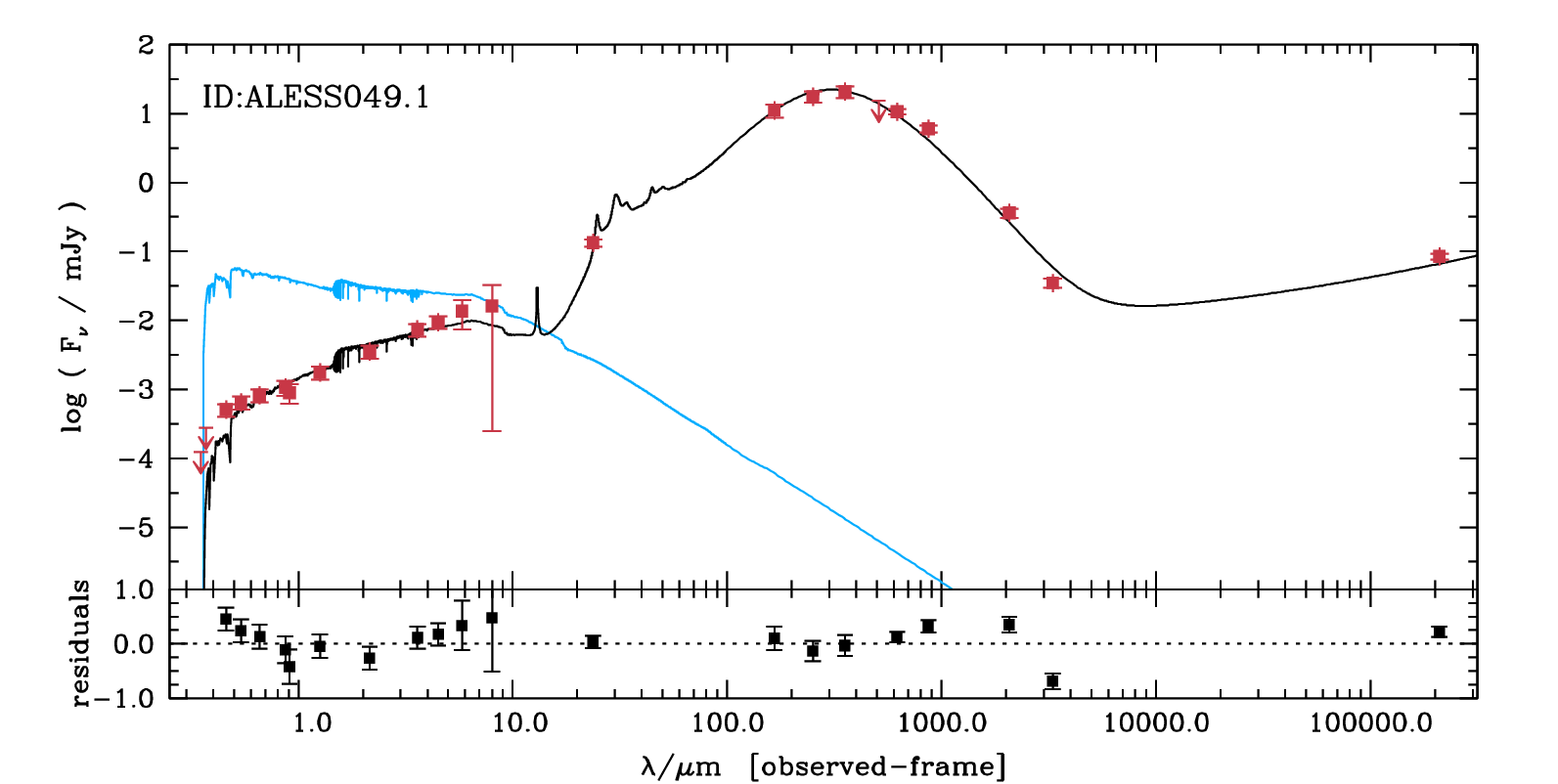}
\includegraphics[width=14cm, clip=true]{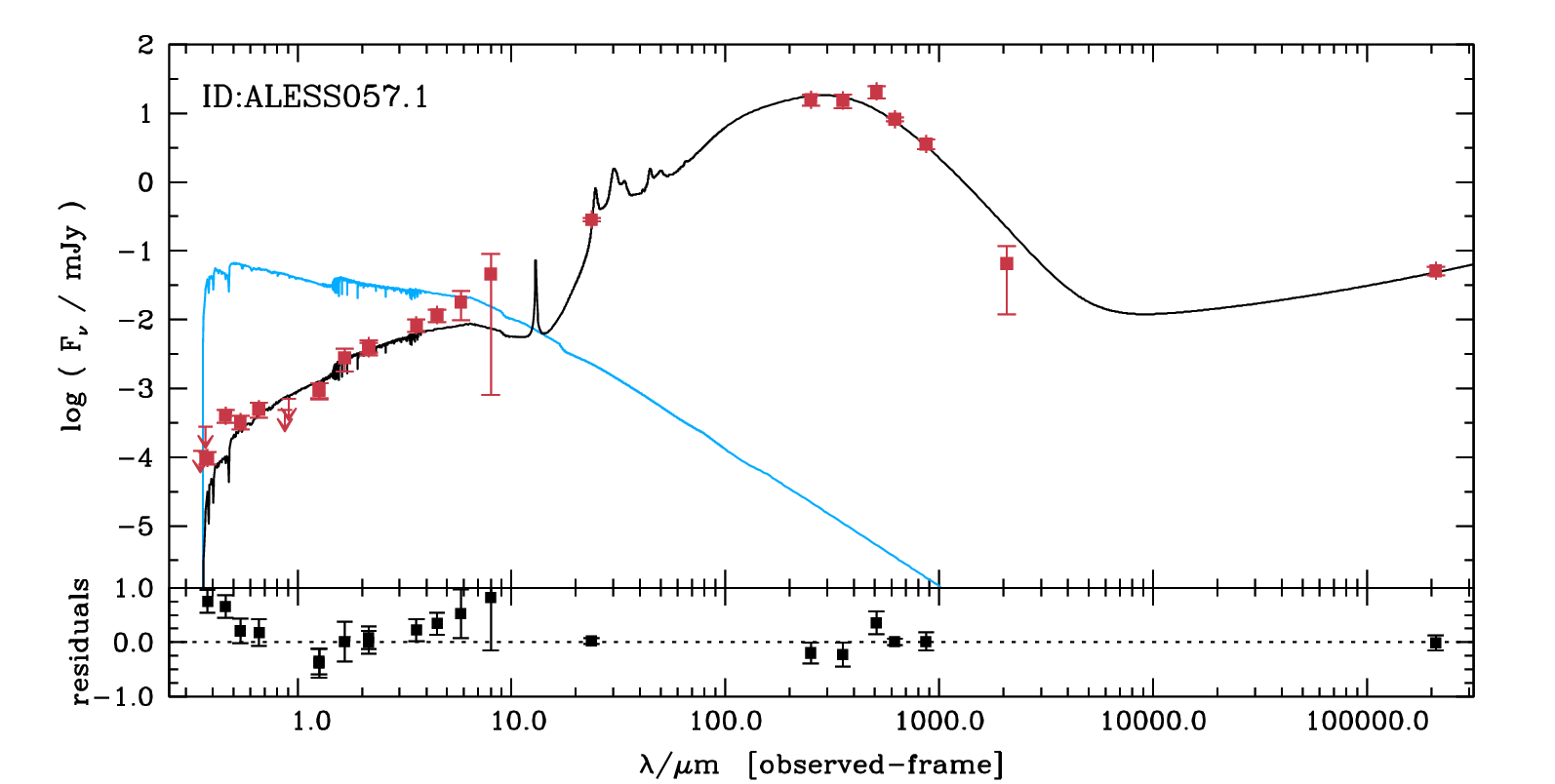}\\

\caption{Multi-wavelength photometry of ALESS~49.1 and 57.1, with best-fitting spectral energy distribution for ALESS~49.1 (\emph{left}) and ALESS~57.1 (\emph{right}) inferred from {\sc MagPhys} modelling (black). The UV-to-radio photometric data are taken from \citet{Swinbank2014} and supplemented by the new ALMA Band~3 \citet{Wardlow2018}, Band~4 (da Cunha et al., in prep.) and Band~8 (this work) continuum observations, which provide improved constraints on the dust thermal emission. The unattenuated stellar spectrum is shown in blue. The reduced $\chi^2$ is $\leq 3 $ for both sources. \label{fig:sed_models}}
\end{centering}
\end{figure*}

\bibliography{ALESS_CII}

\end{document}